\documentclass[12pt,preprint]{aastex}

\usepackage{natbib}
\slugcomment{}

\shorttitle{Pattern Speeds of BIMA-SONG Spirals}
\shortauthors{Rand \& Wallin}

\begin{document}


\title{Pattern Speeds of BIMA-SONG Galaxies with
Molecule-Dominated ISMs Using the Tremaine-Weinberg Method}


\author{Richard J.. Rand}
\affil{Department of Physics and Astronomy, University of New
Mexico, 800 Yale Blvd, NE, Albuquerque, NM 87131}
\email{rjr@phys.unm.edu}
\and

\author{John F. Wallin\altaffilmark{1}}
\affil{Center for Earth Observing and Space Sciences,
Institute for Computational Sciences and Informatics, MS 5c3,
George Mason University, Fairfax, VA 22030}
\email{jwallin@gmu.edu}


\begin{abstract}

We apply the Tremaine-Weinberg method of pattern speed determination to data
cubes of CO emission in six spiral galaxies from the BIMA SONG survey each
with an ISM dominated by molecular gas.  We compare derived pattern speeds
with estimates based on other methods, usually involving the identification of
a predicted behavior at one or more resonances of the pattern(s).  In two
cases (NGC 1068 and NGC 4736) we find evidence for a central bar pattern speed
that is greater than that of the surrounding spiral and roughly consistent
with previous estimates.  However, the spiral pattern speed in both cases is
much larger than previous determinations.  For the barred spirals NGC 3627 and
NGC 4321, the method is insensitive to the bar pattern speed (the bar in each
is nearly parallel to the major axis; in this case the method will not work),
but for the former galaxy the spiral pattern speed found agrees with previous
estimates of the bar pattern speed, suggesting that these two structures are
part of a single pattern.  For the latter, the spiral pattern speed found is
in agreement with several previous determinations.  For the flocculent spiral
NGC 4414 and the ``Evil Eye'' galaxy NGC 4826, the method does not support the
presence of a large-scale coherent pattern.  We also apply the method to a
simulated barred galaxy in order to demonstrate its validity and to understand
its sensitivity to various observational parameters.  In addition, we study
the results of applying the method to a simulated, clumpy axisymmetric disk
with no wave present.  The TW method in this case may falsely indicate
a well-defined pattern.

\end{abstract}


%% subject header.  Objects should be in the appropriate "individual"
%% headers (e.g. quasars: individual, stars: individual, etc.) with the
%% additional provision that the total number of headers, including each
%% individual object, not exceed six.  The \objectname{} macro, and its
%% alias \object{}, is used to mark each object.  The macro takes the object
%% name as its primary argument.  This name will appear in the paper
%% and serve as the link's anchor in the electronic edition if the name
%% is recognized by the data centers.  The macro also takes an optional
%% argument in parentheses in cases where the data center identification
%% differs from what is to be printed in the paper.

\keywords{galaxies: ISM --- galaxies: kinematics and dynamics ---
galaxies: spiral --- 
galaxies: individual(\object{NGC 1068},\object{NGC 3627},\object{NGC 4321},
\object{NGC 4414},\object{NGC 4736},\object{NGC 4826}) --- methods: numerical}


\section{Introduction}

While it is generally accepted that spirals and bars in galaxies are largely
wave phenomena, one of the fundamental difficulties in their study has been
the measurement of the angular speed of the wave patterns, especially for
spirals.  These angular frequencies are not directly observable but crucial
for understanding the rate by which spirals and bars affect the evolution of
galaxies.  Spiral density waves have been shown to trigger star formation in
grand-design spirals (e.g. \citealp*{1993ApJ...410...68R};
\citealp{1992ApJ...385L..37K}), so the rate at which gas encounters spiral
arms may affect the rate of star formation.  Scattering of stars off spiral
arms is likely to be a significant source of disk heating, which may be
particularly important in late-type spirals (\citealp*{2001gddg.conf..221M}).
The rate of heating in this way is therefore linked to the rate at which stars
encounter the arms, and thus the pattern speed.

Traditionally, pattern speeds of spirals have been determined by identifying
theoretically predicted behaviors associated with various resonances of the
wave with the periodic motions of the stars and gas, such as the Lindblad
and Corotation radii (e.g. \citealp{1974ApJS...27..449T};
\citealp*{1989ApJ...343..602E}; \citealp*{1991ApJ...381..130L}).  Knowledge of
the rotation curve is also needed.  Such analyses have yielded pattern speed
estimates for several galaxies.

However, theoretical predictions for resonant behavior vary significantly, not
least because the nonlinear response at resonances is difficult to address
either analytically or by simulation, and is different for stars and gas.  We
do not attempt to present a full summary of the various theoretical studies,
but rather highlight the diversity of predictions.  For instance, the
beginning of spiral structure is often associated with the Inner Lindblad
Resonance (ILR).  However, waves may be sustained beyond the Lindblad
Resonances in the gas (and therefore in the young stars) but not in the old
stellar disk \citep{1994pgsd.conf...35B}.  But although waves in the gas might
propagate through these resonances, it is also possible that gaps in the gas
distribution may be created at these locations, depending on gas viscosity and
other parameters, if the stellar wave is strong enough
\citep{1993PASP..105..664L}.  Of course, depending on the rotation curve and
pattern speed, some galaxies may not have an ILR.  Good knowledge of the
rotation curve is necessary to determine whether a galaxy has an ILR (or two),
and its location.  For a typical rotation curve with a turnover, the ILR is
often near the turnover radius where the slope of the curve, and thus the
epicyclic frequency, is changing rapidly, while noncircular motions can hinder
determination of the axisymmetric rotation curve.  \citet{1993PASP..105..664L}
also notes reasons why the gaseous response at corotation may be complex.

There have also been various theoretical predictions regarding the outer
extent of spiral structure, which have been used to estimate pattern speeds.
For instance, \citet{1970IAUS...38..377L} predicted that spiral waves should
be damped at the Corotation Radius (CR), but attempts to determine the pattern
speed of M81 based on this prediction were sensitive to the depth of the
imaging and the choice of tracer (e.g. stars vs HI), which led to
substantially different results (see \citealt{1998ApJS..115..203W}).  This
prediction was superseded by \citet{1970ApJ...160...89S}, who argued that
significant damping should occur at the Outer Lindblad Resonance (OLR).  Orbit
calculations by \citet{1986A&A...155...11C} indicate on the other hand that
strong spirals can only be maintained up to the 4:1 resonance.

Pattern speeds of some spirals -- M51, for example
(\citealp*{1993A&A...274..148G}) -- have also been estimated by matching
observed spiral morphology with predictions from numerical simulations, while
the change in character of the streaming motions induced by the wave on either
side of corotation has been used to find pattern speeds in other spirals
(e.g. \citealp {1993ApJ...414..487C}; \citealp*{1998ApJ...494L..37E};
\citealp{2003A&A...407..485G}; \citealp{2003A&A...405...89C}).  This method is
potentially very useful as the change in streaming motions is a more direct
and simple prediction of density wave theory, although it may not always be
easy to identify observationally.

While the main focus of this paper is on spiral patterns, we note that for
bars, there is a body of observational and theoretical evidence that the CR
lies just beyond the end of the bar (e.g. \citealp{1980A&A....81..198C};
\citealp{1996AJ....111.2233E}, \citealp*{1997ApJ...476L..73P};
\citealp*{1998AJ....116.2136A}).  Based on this result, many bar pattern
speeds have been estimated [see compilation in \citet{1996AJ....111.2233E}].

Because of the various assumptions involved in these methods of determining
pattern speeds in spiral galaxies, a completely independent method, with
applicability to a large number of galaxies while not relying, for example, on
any particular theory of density waves or a numerical simulation, is
desirable.  In this paper, we continue our exploration of the use of the
Tremaine-Weinberg (\citealp*[hereafter TW]{1984ApJ...282L...5T}) method in
determining pattern speeds in spiral galaxies with molecule-dominated ISMs,
begun in Zimmer, Rand, \& McGraw (2004, hereafter ZRM).

\subsection{The Tremaine-Weinberg Method}

The Tremaine-Weinberg method (hereafter TW method) allows pattern speeds to be
determined for spirals and bars under several assumptions:

1. The disk has a single, well defined pattern speed.
                                                                              
2. A disk component can be found that obeys the continuity
equation as it orbits through the spiral pattern: it is
neither created nor destroyed in significant amounts over an orbit.
                                                                              
3. The relation between the emission from this component and its surface
density is linear, or if not, suspected deviations from linearity can be
modeled.
                                                                              
4. The surface density of the component goes to zero (within noise) at
some radius and all azimuths within the map boundary.

5. The disk of the galaxy must be flat (no warps), at least out to a distance
where the intensity of the tracer is nearly zero.  Motions and variations in
structure perpendicular to the disk must be negligible.

Under these assumptions, TW show how the continuity equation,
written in cartesian coordinates in the plane of the disk:
                                                                              
$${\partial\,\Sigma(x,y,t)\over \partial\,t} + {\partial \over \partial\,x} [\Sigma(x,y,t)\,v_x(x,y,t)] + {\partial \over \partial\,y} [\Sigma(x,y,t)\,v_y(x,y,t)] = 0 \eqno{(1)}$$
                                                                              
can be integrated in the major and minor axis directions to allow the pattern
speed to be determined from maps of the intensity and velocity field of the
tracer (skipping several steps of derivation; see TW):
                                                                              
$$\Omega_p\int_{-\infty}^{\infty}\Sigma(x,y,t)x\,dx = \int_{-\infty}^{\infty}\Sigma(x,y,t)v_y(x,y,t)\,dx \eqno{(2)}$$

where $\Omega_p$ is the pattern speed, $\Sigma$ is the surface density of the
component, $x$ and $y$ are the major and minor axis coordinates, and $v_x$
and $v_y$ are the velocity components along those axes -- thus $v_y$ is simply
the observed velocity divided by the galaxy inclination.  The pattern speed
enters because the time derivative of the surface density at a given location
depends on the rate at which the rotating pattern carries material through that
location (in a stationary spiral, the pattern speed would be zero, and the
surface density at every point would not change with time).

Equation (2) says that if the relation between the emission and surface
density of the component is understood, then $\Omega_p$, modulo the galaxy
inclination, can be determined by dividing the sum of the intensity-weighted
observed velocity along a line parallel to the major axis by the
intensity-weighted position coordinate.  Since each line parallel to the major
axis provides an independent measurement of $\Omega_p$, many determinations
can be made for a galaxy, depending on resolution and sensitivity.  A better
value can then be found by averaging measurements of $\Omega_p$ for individual
apertures.  If more than one pattern is present for a given aperture [as in,
e.g., the bar-within-bar models of \citet{1990ApJ...363..391P} and
\citet{1993A&A...277...27F}, or the bar-within-spiral models of
\citet{1988MNRAS.231P..25S}], $\Omega_p$ will have contributions from the
various patterns.  Alternatively, one can plot the intensity-weighted,
inclination-corrected line-of-sight velocity, hereafter denoted $<$$v$$>$,
vs. the intensity-weighted position coordinate, $<$$x$$>$, for all apertures
and find the best-fit slope -- a variation introduced by
\citet{1995MNRAS.274..933M}.  This refinement has the advantages that errors
in the systemic velocity and $y$-coordinate of the kinematic center of the
galaxy do not affect the final value.  Also, apertures close to the center of
the galaxy, for which values of $<$$v$$>$ and $<$$x$$>$ tend to be small,
leading to noisy measurements of $\Omega_p$, carry less weight in the fit.
However, the best fit slope will be biased by the outer apertures.

The TW method has been mainly applied to early-type bars using starlight and
absorption-line kinematics from long-slit spectra [see
\citet*{2003MNRAS.345..261G} and references therein], since older stars will
survive many passages through the pattern and the extinction problem, which is
of concern because of the third assumption above, is minimized.  Modified
forms of the method have also been used by \citet{2000ApJ...539L..17S} to
determine the pattern speed of the nucleus of M31, and by
\citet*{2002MNRAS.334..355D}, who used 1612 MHz emission from OH/IR
stars in the Milky Way to determine the speed of the pattern (or patterns)
they participate in.

Applications to gaseous phases are complicated especially by the second and
third assumptions above.  Phase changes as gas orbits through a pattern will
lead to the second assumption being violated in general if only one gaseous
phase is studied.  In general, the ISM mass will be dominated by the molecular
and atomic phases, and there may be significant exchange of gas from one phase
to the other as a result of molecule formation in self-shielding atomic gas
clouds and dissociation of molecular gas clouds by stellar radiation,
especially in spiral arms.  However, if the ISM is dominated everywhere by
either atomic or molecular gas, then it is clear that phase changes involve
only a small percentage of the dominant phase.  Also, star formation is a
potential sink in the continuity equation, but the process is inefficient.
\citet{1998ARA&A..36..189K} finds that the average spiral converts only 4.8\%
of its gas into stars every 10$^8$ years.  CO emission also has the advantage
that it is largely concentrated to the inner disks of galaxies where warps are
less likely to occur.  While 21-cm emission provides a reliable measure of HI
column densities, the standard tracer of molecular column densities, CO
emission, is subject to larger uncertainties, particularly in metallicity.
The TW method has been applied to the spiral M81 by
\citet{1998ApJS..115..203W} and the Blue Compact Dwarf NGC 2915 by
\citet{1999AJ....118.2158B} using 21-cm observations.

Application of the method to CO observations in three galaxies with
molecule-dominated ISMs has been carried out by ZRM and this paper extends
that work.  Pattern speeds were measured for M51, M83, and NGC 6946, using
fully-sampled CO 1--0 maps from the literature, with results consistent with
determinations based on behavior at resonances for M51
\citep{1974ApJS...27..449T,1989ApJ...343..602E} and M83
\citep{1991ApJ...381..130L}.  The measured speed for M51, $38 \pm 7$ km
s$^{-1}$ kpc$^{-1}$, is however, inconsistent with the value of 27 km s$^{-1}$
kpc$^{-1}$ from \citet{1993A&A...274..148G}, determined from a comparison of
the observed CO morphology with a simulation of the dynamics of the molecular
clouds.  ZRM also found evidence in M51 for a faster pattern speed for the
central bar.

ZRM also tested the method for sensitivity to a variation in the conversion
factor between CO intensity and H$_2$ column density (commonly denoted by $X$)
with metallicity, evidence for which is discussed by
\citet*{2002A&A...384...33B},
\citet{1997A&A...325..124D},\citet*{1996PASJ...48..275A}, and
\citet{1995ApJ...448L..97W}, with recent studies suggesting a linear
relationship.  Such an assumed variation of $X$ (using measured metallicity
gradients) in these galaxies produced no significant change in the derived
pattern speed.  There are also suggestions that $X$ may vary between arm and
interarm regions in M51 by perhaps a factor of 2 \citep{1993A&A...274..148G},
but such a variation also produced no significant change to the derived
pattern speed.  Note that the overall scaling of $X$ for a given galaxy does
not affect the TW method, although a value must be assumed to justify
molecular dominance of the ISM.  In this paper, we assume by default a value,
appropriate for solar metallicity, of $X=2 \times 10^{20}$ mol cm$^{-2}$ (K km
s$^{-1}$)$^{-1}$, consistent with the relationship of $X$ with metallicity
found by \citet{2002A&A...384...33B}.  ZRM also showed that the maps used
easily extend far enough to allow the ratio of the integrals that are used
to calculate $<$$v$$>$ and $<$$x$$>$ to converge to stable values for most
apertures in each galaxy, and to allow the slope in the plot of $<$$v$$>$ vs.
$<$$x$$>$ to converge to a well-defined value.

A great advantage to using data cubes over long-slit spectra is that the
quantities $<$$v$$>$ and $<$$x$$>$ are readily derived from moment maps, and
the sensitivity to errors in the assumed position angle (PA) of the major
axis, the severity of which has been demonstrated by
\citet{2003MNRAS.342.1194D}, can be directly assessed by varying this
parameter.

In this paper, we extend the ZRM study to a larger sample of galaxies, which
we argue all have molecule-dominated ISMs, from the BIMA Survey of Nearby
Galaxies (BIMA SONG; \citealt{2003ApJS..145..259H}).  These 44 galaxies
encompass all spirals (except M33) between Sa and Sd with a redshift less than
2000 km s$^{-1}$, inclination$<$ 70$\degr$, Dec. $>$ --20$\degr$, and $B<11$
mag.  Most were observed with a seven-pointing mosaic covering 200'' on a
side.  Of these galaxies (excluding M51 and NGC 6946), 22 have single dish
data from the NRAO 12-m telescope included in the final maps so that there is
no missing flux problem common to interferometers.  They all resolve spiral
and bar structure (typically 6'' resolution), although in some galaxies only a
bar is well detected.  Fifteen of the galaxies with single-dish CO data have
high-resolution Very Large Array (VLA) or Westerbork Synthesis Radio Telescope
HI data with in the literature.  Six of these have global H$_2$/HI mass ratios
greater than 0.9 \citep{1989ApJ...347L..55Y, 1993A&A...272..123S}.  These are
the subject of this paper.  Further papers will explore galaxies with
HI-dominated ISMs and those for which neither phase dominates (for the latter,
molecular and atomic gas must be considered together in the continuity
equation).  Properties of these galaxies are shown in Table 1.

First, however, we carry out some explorations of the TW method.  We first
show the results of applying the method to a clumpy disk with no wave as a
cautionary example.  Also, although our main focus is on spiral patterns, we
demonstrate the effectiveness of the method and its robustness to various
parameters using simulations of a bar.

\section{Tests of the Tremaine-Weinberg Method}

\subsection{A Clumpy Axisymmetric Disk}

One of the main assumptions of the TW method is that the galaxy have a
single, clear wave pattern.  However, as we show in this subsection, even in
the absence of any wave, use of the TW method may still lead to the
erroneous conclusion that a wave pattern with a well-defined speed exists.

We use the Groningen Image Processing System (GIPSY) program GALMOD to
generate an axisymmetric, clumpy disk of gas with a specified (constant apart
from the clumpiness) surface density, rotation curve, inclination and major
axis PA.  We place the model galaxy at a distance of 10 Mpc, with a pixel size
of 1''.  We smooth the simulated data cube to 6'' resolution, and apply the TW
method to zeroth- and first-moment maps created with no intensity cutoffs,
using apertures spaced by 6''.  The degree of clumpiness results in a
root-mean-square variation of 2\% of the mean value in the zeroth-moment map.
Since the cube is built by counting clouds in spatial and velocity bins,
generated by probability distribution functions using Monte Carlo techniques,
the noise is essentially governed by Poisson statistics.

\begin{figure}
\epsscale{.80}
\plotone{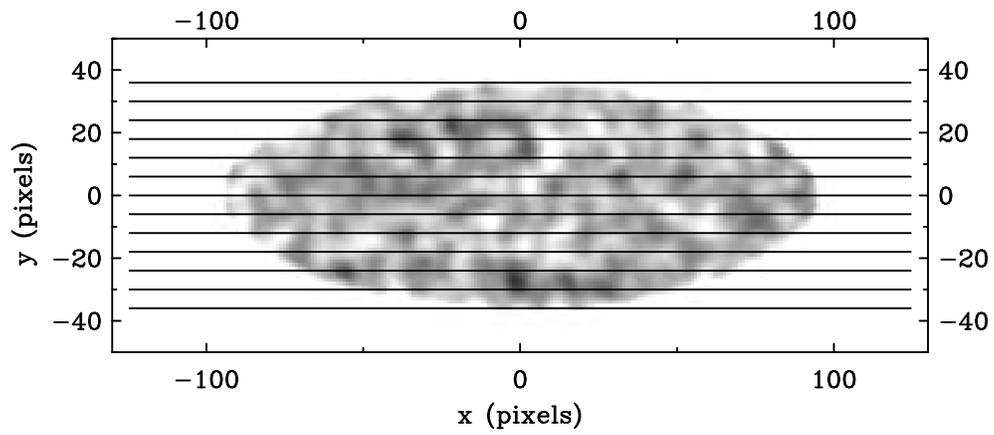}
\caption{Zeroth-moment map of a modeled clumpy, axisymmetric disk inclined
at 65$\degr$, showing the apertures used in the TW method.  Aperture 1
is southernmost.\label{fig1}}
\end{figure}

\begin{figure}
\epsscale{.80}
\plotone{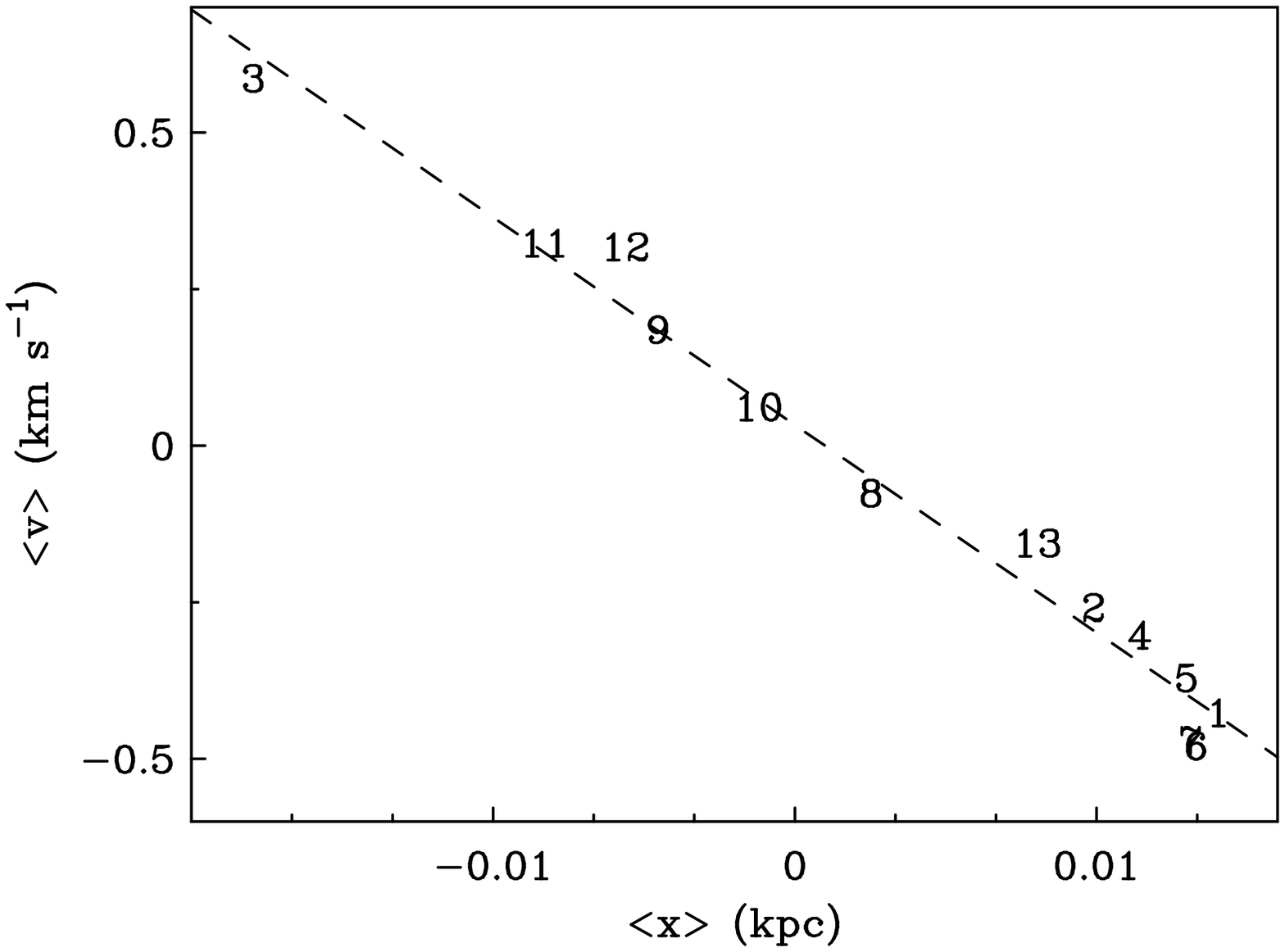}
\caption{Plot of $<$$v$$>$ vs. $<$$x$$>$ for the apertures shown in Figure
1.  The dotted line is the best-fit straight line to all apertures.
\label{fig2}}
\end{figure}

\begin{figure}
\epsscale{.80}
\plotone{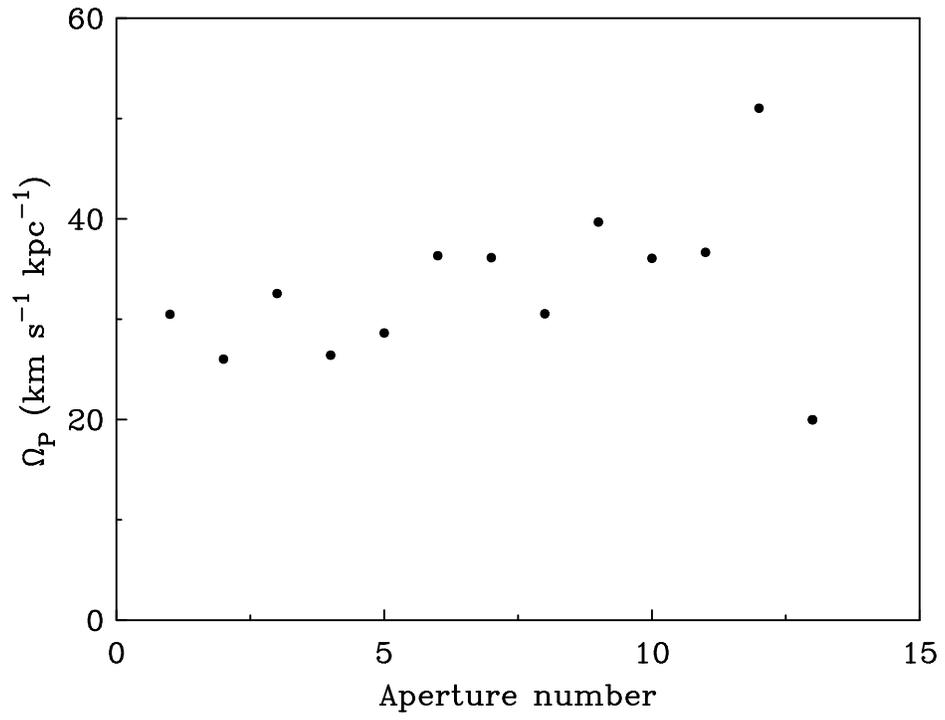}
\caption{Plot of pattern speed vs. aperture number for the apertures
shown in Figure 1.\label{fig3}}
\end{figure}

\begin{figure}
\epsscale{.80}
\plotone{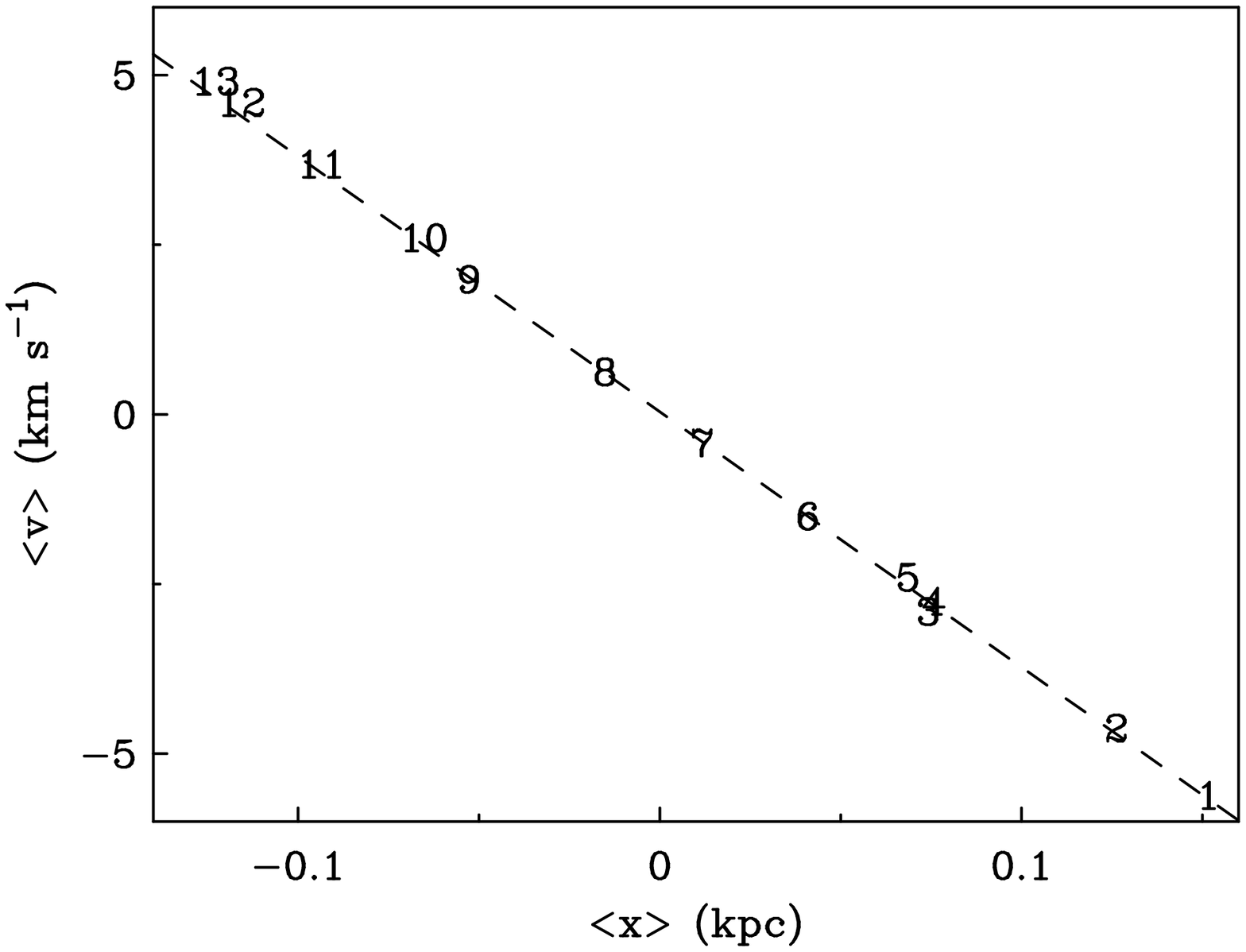}
\caption{Plot of $<$$v$$>$ vs. $<$$x$$>$ for the clumpy,
axisymmetric disk shown in Figure 1, but using an incorrect PA of 91$\degr$.
The dotted line is the best-fit straight line to all apertures.
\label{fig4}}
\end{figure}

The first case we study is a galaxy with a rising rotation curve with a
gradient of 31 km s$^{-1}$ kpc$^{-1}$ at an inclination of 65$\degr$.  Figure
1 shows a zeroth-moment map and the apertures used in the TW method.  The
root-mean-square scatter in the zeroth-moment map is 72\% of the mean value.
Figure 2 shows the resulting plot of $<$$v$$>$ vs. $<$$x$$>$ (hereafter
$<$$v$$>$--$<$$x$$>$ plot), and Figure 3 shows pattern speed vs. aperture
number.  There is a clear correlation in Figure 2, and a well defined mean
value in Figure 3, suggesting that a pattern is present.  However, this is an
artifact of the clumpiness.  First, note that negative values of $x$
correspond to the receding side of the galaxy (this convention is true
throughout this paper -- as a consequence, we will quote absolute values of
all best fit slopes).  Then, along a given aperture, if the distribution of
clumps biases $<$$x$$>$ toward a positive (or negative) value, then they will
also bias $<$$v$$>$ toward a negative (positive) value.  However, the range of
values of $<$$v$$>$ and $<$$x$$>$ will depend on the degree of clumpiness, and
is here very small.  It is also easy to show in general for a linearly rising
rotation curve that the slope in the $<$$v$$>$--$<$$x$$>$ plot will tend to
the slope of the rotation curve.  In this case, the best fit slope is $33.1
\pm 1.4$ km s$^{-1}$ kpc$^{-1}$ (the error bar is derived from the scatter in
the plot).  Another important indicator that no pattern is present is in the
sequence of aperture numbers in the $<$$v$$>$--$<$$x$$>$ plot.  Subsets of
aperture numbers only form sequences when they are dominated by a clump or a
group of clumps crossing several apertures (artificial sequencing can also be
produced if the apertures oversample the resolution).  Otherwise, there should
be no sequence.  For random clumpiness, a given aperture is just as likely to
have a positive value of $<$$x$$>$ or $<$$v$$>$ as a negative value.

For a more typical spiral with a rotation curve turning over to a flat value
at some radius, the slope in the $<$$v$$>$--$<$$x$$>$ plot should be less than
the gradient of the rising part of the curve, depending on the distribution of
clumps and the exact shape of the curve.  Indeed, for a rotation curve that
rises with the above gradient in the inner half of the disk before becoming
flat, we measure a slope in the $<$$v$$>$--$<$$x$$>$ plot of $20.9 \pm 1.2$ km
s$^{-1}$ kpc$^{-1}$.

The results of this simulation are very sensitive to the assumed PA.  The
relative lack of a sequence of aperture numbers can vanish if the position
angle is inaccurately known.  Figure 4 shows $<$$v$$>$ vs. $<$$x$$>$ for a
$+1\degr$ error in PA.  The aperture numbers are now in sequence, the scatter
is reduced, and the range of $<$$v$$>$ and $<$$x$$>$ is larger.  These are
natural consequences of the asymmetry introduced by this error (see \citealt{
2003MNRAS.342.1194D}).  For a $-1\degr$ error, the plot is very similar but
with the aperture sequence essentially reversed.  The slope in both cases is
$37.7 \pm 0.2$ km s$^{-1}$ kpc$^{-1}$.  In general, PA errors increase the
slope.

These effects are less obvious if there is more scatter, which will occur if
the disk is clumpier or the inclination is reduced.  For example, for
$i=15\degr$, the slope is $31.2 \pm 2.8$ km s$^{-1}$ kpc$^{-1}$, with even
less evidence for sequencing of aperture numbers and a larger range of
$<$$v$$>$ and $<$$x$$>$.  PA errors of $+/- 1\degr$ still introduce clear
sequencing.  For a given inclination, we find that decreasing the degree of
clumpiness does not significantly change the slope but does decrease the range
of $<$$v$$>$ and $<$$x$$>$, as expected.  For a perfectly smooth disk, all
apertures would yield values of zero for $<$$v$$>$ and $<$$x$$>$.

\subsection{The Method Applied to a Simulated Bar}

Here we carry out tests of the TW method on numerical simulations of a bar, in
order to demonstrate the utility of the method and to understand its
sensitivity to the angle of the bar relative to the major and minor axes, the
angular resolution, the inclination, and [as already explored by
\citet{2003MNRAS.342.1194D}] the uncertainty in the PA.  A bar presents more
constraints than a spiral because the latter has no preferred position angle
relative to the principle axes (as long as it wraps through a large enough
angle).

To create a bar with a defined pattern speed, we created a self-consistent
three-component model of a galaxy following the commonly used general recipe
of \citet{1993ApJS...86..389H}.  His procedure includes parameters for a disk,
bulge, and a halo.  We changed half the disk particles into gas particles
using smoothed particle hydrodynamics (SPH).
 
The parameters for the galaxy in our simulations are given in Table 2.
Obviously the masses listed are in relative units, and can be rescaled as
needed.  The primary difference between these parameters and those given by
Hernquist are the ratios between disk and halo.  By setting the mass ratio
between the disk and halo to be $1:1$, we create a system that is
dynamically unstable to bar formation as shown by \citet{1973ApJ...186..467O}
and others.

The code used to evolve the system is ``mass99'', a tree gravity + SPH code
developed at George Mason University.  For these simulations, the code was set
to have similar run parameters to those in other galaxy simulations
\citep{1989ApJS...70..419H}.  There are 32,000 total particles, of which 6400
are SPH particles.  The simulation grid is 500 pixels on a side.  In physical
units, the masses in Table 2 are in units of $2.8 \times 10^{10}$ M$\sun$, and
the pixel size is 120 pc.

We choose a frame from the simulation where the bar is clearly well developed,
is not significantly evolving in shape, and has settled into a constant
angular speed, which we measure directly from an animation of the time
evolution of the galaxy to be $22 \pm 2$ km s$^{-1}$ kpc$^{-1}$.  This angular
speed was checked over several periods to make sure that it was not rapidly
evolving.

Using the SPH data from the simulations, we created simulated observations of
the gas distribution in our system.  After picking a viewing angle, data were
binned into a three-dimensional grid of projected position and line-of-sight
velocity.  These data were then convolved with a Gaussian filter of a
specified size in each velocity plane and an initial cutoff threshold was
included to drop pixels that were below a specified ``noise'' threshold.
Using the convolved and clipped data, we formed zeroth- and first-moment maps
of the gas distribution.  Phase changes in the gas component were not
modeled.
 
In the frame we used for our primary analysis, a weak spiral exists with the
same pattern speed, along with clumpy disk gas with no pattern.  We found that
further intensity clipping to remove disk gas improves many of the results
described below by decreasing the scatter induced by this clumpiness (the
intensity contour at which the data were clipped is indicated in the relevant
figures).  However, clipping may also remove gas participating in the bar
pattern, violating the assumptions of the TW method, and we find that clipping
at too high an intensity leads to incorrect pattern speed determinations.

As a representative case, we examine a frame from the simulation, viewed at
$i=35\degr$, where the bar makes a large angle with both axes.  Figure 5 shows
the zeroth-moment map and the apertures used, and the first-moment map.  Here
the data cube was convolved to 720 pc resolution before the moment maps were
made.  The clear sign of bar orbits, namely isovelocity contours running at an
angle to the galaxy major axis, is evident.  In all cases the apertures are
spaced by the resolution FWHM.  The contour indicates the clipping level.  The
$<$$v$$>$--$<$$x$$>$ plot is shown in Figure 6, and the best fit slope is $23
\pm 1$ km s$^{-1}$ kpc$^{-1}$, within the error bars of the true value (note
that $<$$x$$>$ in Figure 6 refers to the major axis coordinate, which is
actually the vertical axis in Figure 5 and thus labeled '{\it y}').  Here, the
sequence of aperture numbers reflects the structure of the simulated galaxy:
as one proceeds from aperture 1 to 9, the bulk of the emission steadily moves
from the negative to the positive side of the minor axis.  The sequencing is
slightly different from that of the simulated bar of
\citet{2003MNRAS.342.1194D} because the shape of the bar changes with distance
from its center.  The relatively round iso-density contours of the emission
(his Figure 3) that dominates the first (and last) two apertures results in
values of $<$$v$$>$ and $<$$x$$>$ in his Figure 5 somewhat closer to zero than
the following (preceding) two apertures, which cross the bar where it is
rather elongated.

\begin{figure}

\includegraphics[scale=1,viewport=50 0 574 468]{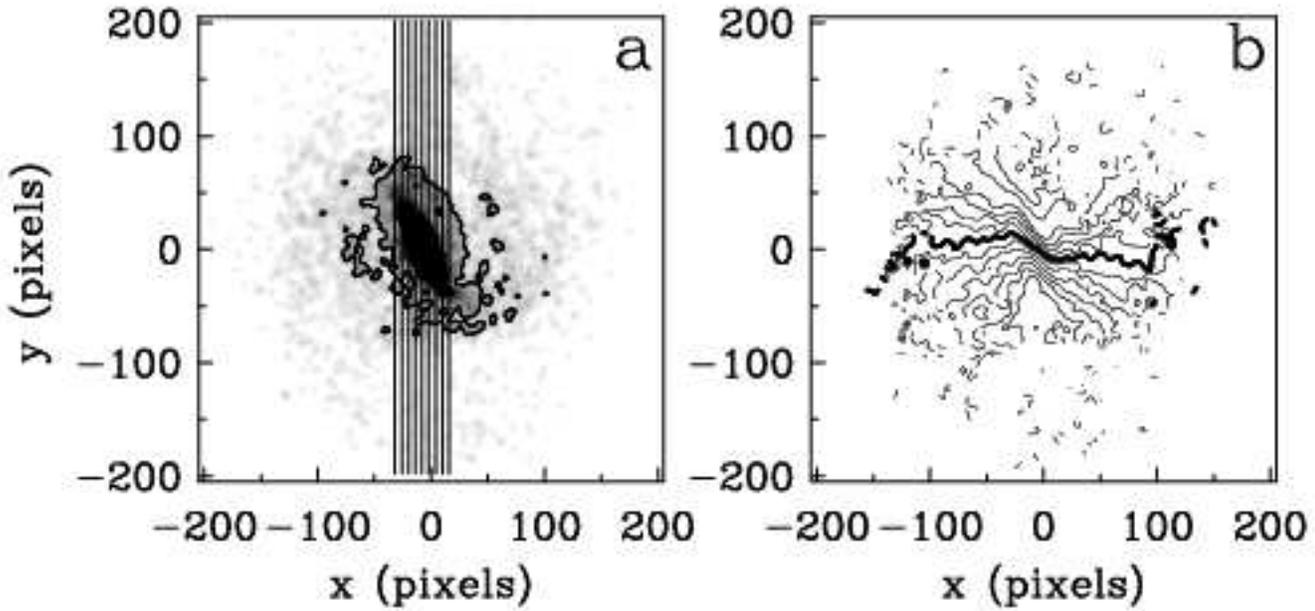}
\caption{In a) is shown the zeroth-moment map of a simulated bar and the
apertures used in the TW method.  The pixel size is 120 pc.  The contour
indicates the level at which the data were clipped to remove the clumpy disk.
In b) is shown the first-moment map.  The systemic velocity is shown by the
heavy contour.  The contour spacing is 20 km s$^{-1}$ with velocities
increasing from north to south.
\label{fig5}}
\end{figure}

\begin{figure}
\epsscale{.80}
\plotone{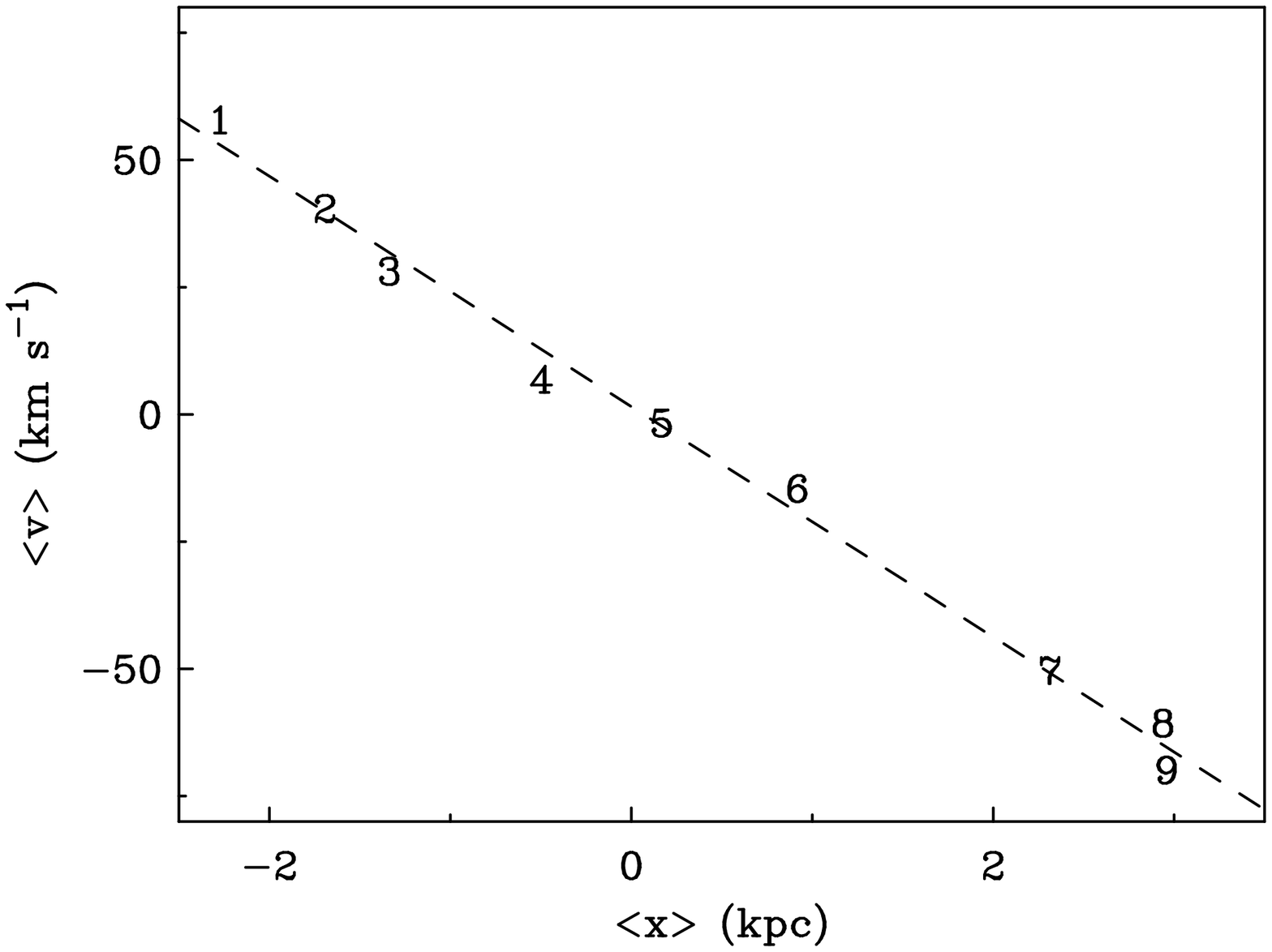}
\caption{Plot of $<$$v$$>$ vs. $<$$x$$>$ for the apertures shown in Figure
5a.  The dotted line is the best-fit straight line to all apertures.\label{fig6}}
\end{figure}

\citet{2003MNRAS.342.1194D} has demonstrated the sensitivity of the TW method
to errors in the assumed major axis position angle.  We confirm this
sensitivity here, finding that, for the above simulation, $+5\degr$ and
$-5\degr$ errors in the PA, typical of real observations, result in best fit
slopes of $15 \pm 1$ km s$^{-1}$ kpc$^{-1}$ and $28 \pm 1$ km s$^{-1}$
kpc$^{-1}$, or percent errors of 35\% and 22\%, respectively.

To study the sensitivity of the TW method on bar viewing parameters, we tilt
the galaxy at four different inclinations (10$\degr$, 35$\degr$, 65$\degr$,
and 80$\degr$) around a range of PAs in increments of 10$\degr$ (sometimes
5$\degr$ when the bar is nearly aligned with a principle axis).  The linear
resolution is again 720 pc.  Results for the derived pattern speed vs. these
angles are shown in Figure 7.  Alignments with the principle axes are
indicated, as well as the pattern speed measured from the animation.  For
$i=80\degr$ and PA$<10\degr$, only three apertures cross the bar, making the
resultant slope highly uncertain, and thus no data points are plotted.  In
general, a pattern speed accurate to within the combined errors can be
measured as long as the bar is not within about $20\degr$ of a principle axis,
for all inclinations, even as high as $80\degr$.  The reason for the failure
of the TW method for bars closely aligned with a principle axis is that,
because of the symmetry introduced in positions and velocities around the
principle axes, both integrals in Eq. 2 tend to zero.  In some cases, a
reliable pattern speed can be measured when the bar is as close as 5$\degr$ to
the major axis, probably due to gas from the beginnings of the spiral arms
being included in many of the apertures.

\begin{figure}
\includegraphics[scale=1,viewport=50 100 574 568]{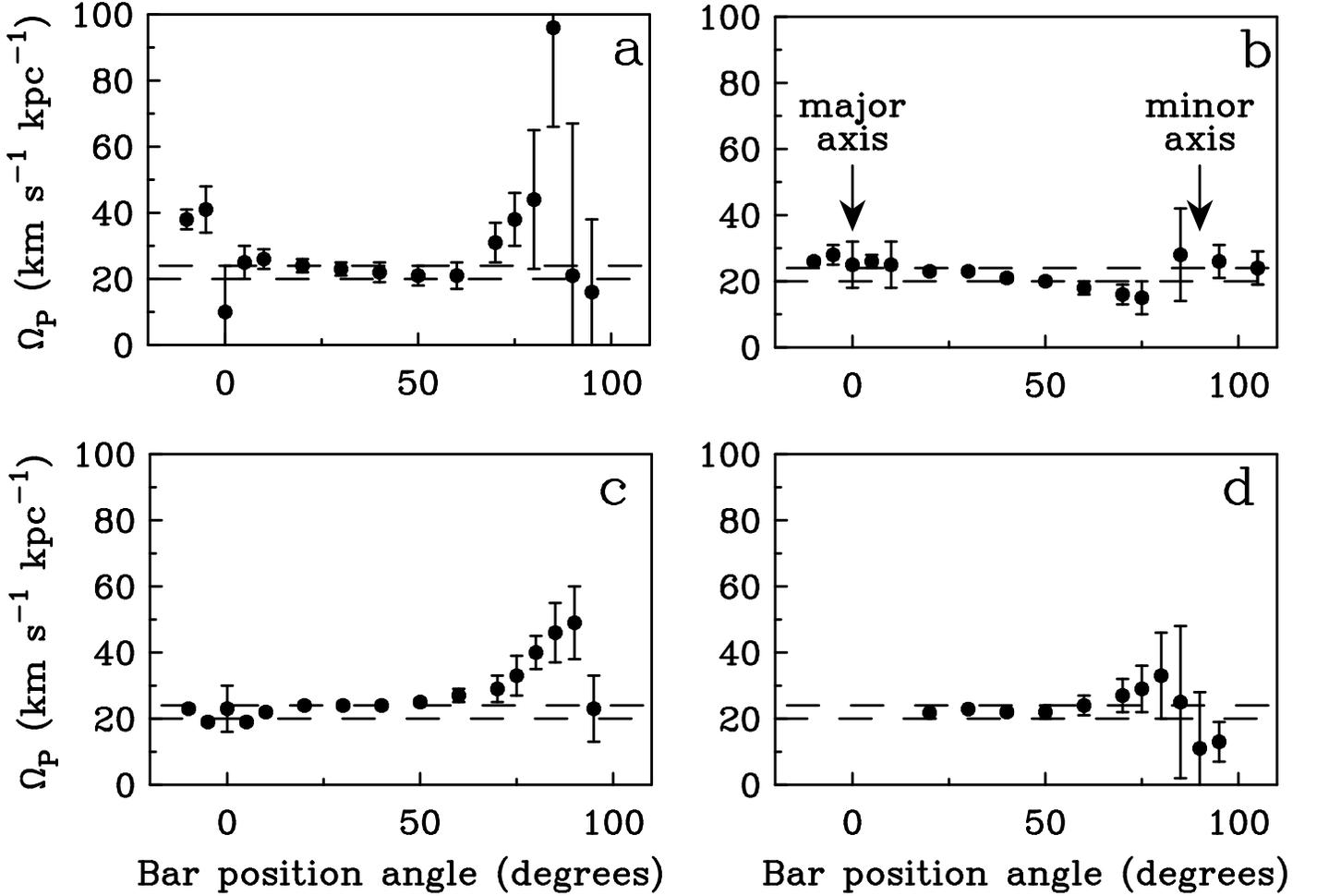}
\caption{Derived pattern speed vs. angle of the bar from the major axis for
four different inclinations: a) 10$\degr$; b) 35$\degr$; c) 65$\degr$; and d)
80$\degr$.  The dashed lines indicate the pattern speed as estimated from an
animation of the time evolution of the simulation of $20-24$ km s$^{-1}$
kpc$^{-1}$.
\label{fig7}}
\end{figure}

\begin{figure}
\includegraphics[scale=1,viewport=50 0 574 568]{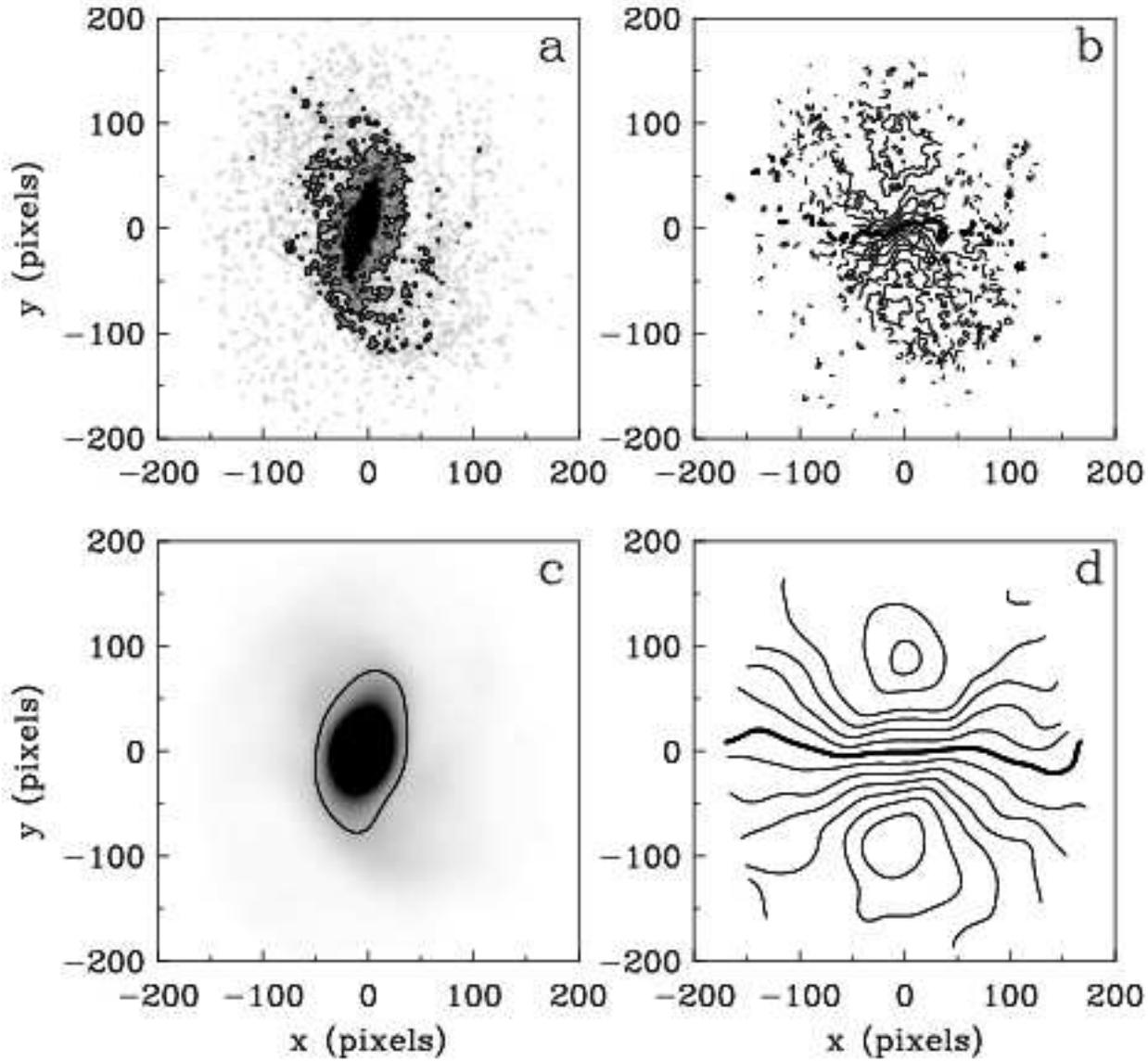}
\caption{In a) and b) are shown zeroth- and first-moment maps, respectively,
of a simulated bar at a resolution of four pixels (480 pc).  In c) and d) are
shown zeroth- and first-moment maps of the same bar at a resolution of 32
pixels (3.8 kpc).  Contours are as described in the caption to Figure 5.
\label{fig8}}
\end{figure}

\begin{figure}
\epsscale{.80}
\plotone{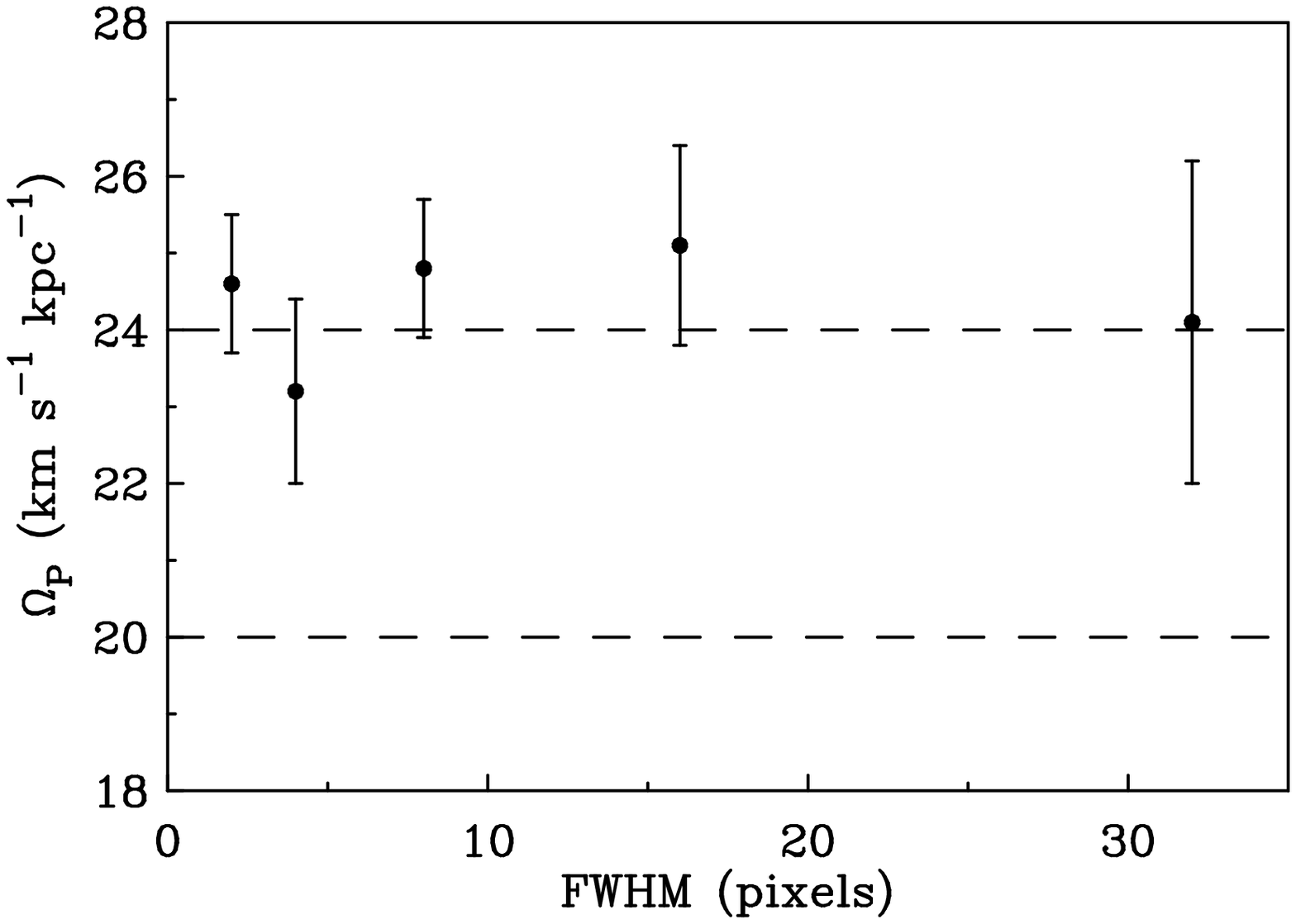}
\caption{Pattern speed vs. Gaussian smoothing FWHM for the simulated bar.
One pixel equals 120 pc.
\label{fig9}}
\end{figure}

Finally, we test the sensitivity of the method to angular resolution.  We
smooth the simulated data cube using Gaussians of FWHM approximately 2, 4, 8,
16 and 32 pixels, before creating moment maps.  As examples, the zeroth- and
first-moment maps at 4- and 32- pixel resolution are shown in Figure 8.  Even
with the highest degree of smoothing, the twist in the
isovelocity contours is still clear.  Figure 9 shows a plot of
best fit slope vs. resolution.  The slopes are consistent within the error
bars (they are slightly larger than the value derived from the animation
because in the chosen frame the bar is fairly well aligned with the major axis).
The implication is that the TW method can be applied to even poor angular
resolution observations as long as the morphological and kinematic signatures
of the pattern are evident.  In ZRM, we performed a similar test by degrading
the original 16'' resolution of the CO data cube of M51 by factors of 2 and 3,
finding no significant difference in the determined spiral pattern speed.  TW
also noted that resolution did not cause a bias in the determination.

%% The \notetoeditor{TEXT} command allows the author to communicate
%% information to the copy editor.  This information will appear as a
%% footnote on the printed copy for the manuscript style file.  Nothing will
%% appear on the printed copy if the preprint or
%% preprint2 style files are used.

\section{Results for the BIMA-SONG Galaxies}

For each galaxy, zeroth- and first-moment maps were made from full-resolution
data cubes.  Angular resolutions are of order 6'' (more precise beam
parameters are given below).  In some galaxies (noted below) with
well-resolved spiral structure, cubes Gaussian-smoothed to 12'' resolution
were used.  In all cases, sensitivity was greatly improved by using the GIPSY
program BLOT to remove noise that can be distinguished as emission features
that clearly do not follow the rotation of the galaxy.  Moment maps were made
using 1--2$\sigma$ cutoffs and excluding any channels at either end of the
cube that were judged not to include real emission.  The quantities $<$$v$$>$
and $<$$x$$>$ were calculated for apertures spaced by approximately one beam
width in all cases.  For all galaxies, the channel width is 10 km s$^{-1}$.
All previous pattern speed determinations quoted are scaled to our adopted
galaxy distances.

\subsection{NGC 1068}

NGC 1068 is a very well studied Seyfert 2 galaxy and is classified as
(R)SA(rs)b in the RC3 catalog \citep{1991trcb.book.....D}.  Running through
the Seyfert nucleus is a 2.3 kpc long [for an assumed distance of 14.4 Mpc;
\citet{1997Ap&SS.248....9B}] stellar bar \citep{1988ApJ...327L..61S,
1989ApJ...343..158T}, also detected in CO (e.g. \citealt{2000ApJ...533..850S}).
A two-arm spiral begins near the ends of the bar, the bright inner segments
wrapped tightly enough to appear as a pseudo-ring at $R \approx 14-20''$.  The
bar and spiral are clearly evident in the zeroth-moment map from the BIMA SONG
data (Figure 10).  Note that the bar runs at a substantial angle to both
principle axes.

\begin{figure}
\epsscale{.80}
\plotone{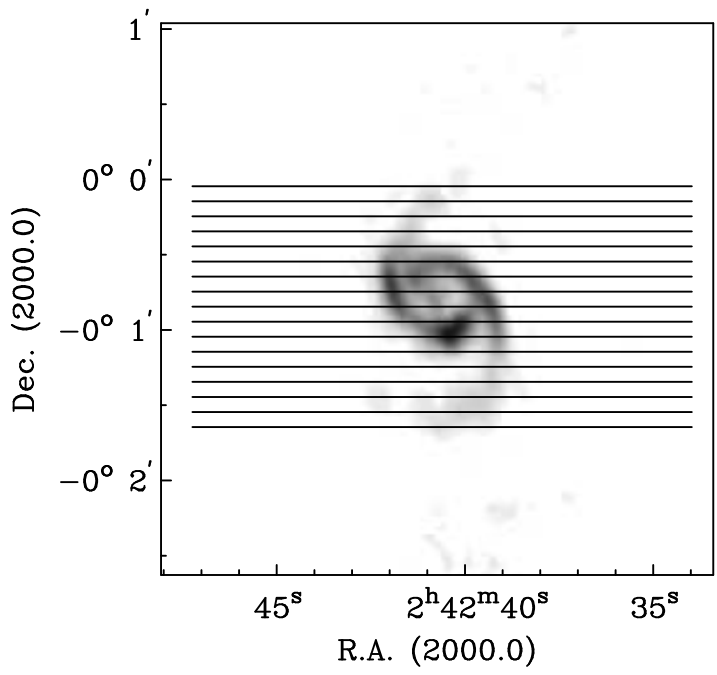}
\caption{Zeroth-moment map for NGC 1068.  Apertures used in the TW
calculation are shown, with the southernmost being Aperture 1.
\label{fig10}}
\end{figure}

\begin{figure}
\epsscale{.80}
\plotone{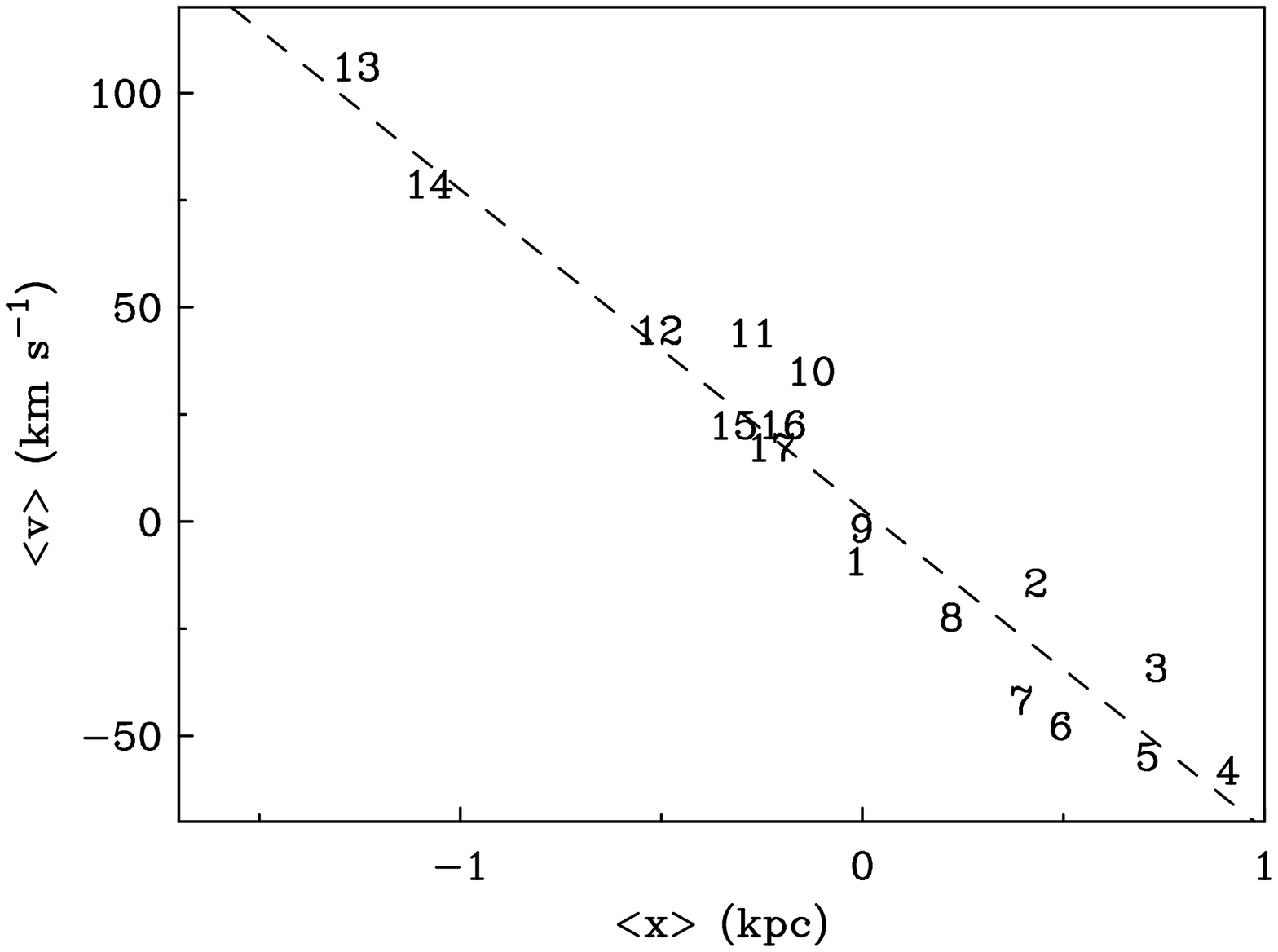}
\caption{Plot of $<$$v$$>$ vs. $<$$x$$>$ for the apertures shown in Figure 10.
The dotted line is the best-fit straight line to all apertures.\label{fig11}}
\end{figure}

On the basis of $BRI$ images analyzed by \citet{1997Ap&SS.248....9B},
\citet{2000ApJ...533..850S} conclude that there is a large-scale bar of radius
$\approx 120''$ oriented nearly along the minor axis.  By assuming corotation
is near the end of this bar, they estimate a pattern speed of $20 \pm 10$ km
s$^{-1}$ kpc$^{-1}$.  They interpret the pseudo-ring as a pile-up of gas
between the outer ILR (oILR) and inner ILR (iILR).  They further interpret the
CR of the inner bar to coincide with the oILR of the outer bar,
yielding a pattern speed for the former of $\sim 140$ km s$^{-1}$ kpc$^{-1}$.
\citet{1995ApJ...450...90H} derive a slightly larger speed of $160$ km
s$^{-1}$ kpc$^{-1}$, using a slightly shorter measured inner bar length of
15'' and assuming the pseudo-ring lies between corotation and the OLR of the
inner bar (they find that the pseudo-ring molecular gas shows streaming
motions indicative of a location beyond the CR). 

The inclination of NGC 1068 is taken to be 40$\degr$
(e.g. \citealt{2000ApJ...533..850S}).  The major axis PA is a matter of some
uncertainty.  From HI, at radii between 30'' and 70'', where most of the CO
emission is concentrated, \citet{1997Ap&SS.248...23B} find a PA of $271\degr
\pm 3\degr$.  From OVRO CO 1--0 observations, \citet{2000ApJ...533..850S} find
that the PA is about $280\degr$ outside $R=15''$, increasing within 10'' of
the nucleus to about 305$\degr$.  \citet{1995ApJ...450...90H}, using BIMA CO
1--0 data, found a PA of 264$\degr$ in the region of the spiral arms, with a
similar increase within 10'' of the center as found by
\citet{2000ApJ...533..850S}.  It is clear from Figure 8 of
\citet{2003ApJS..145..259H}, showing the BIMA SONG CO 1--0 velocity field, or
Figure 3 of \citet{1997Ap&SS.248...33D}, showing the H$\alpha$ velocity field,
that the kinematic major and minor axes are not parallel, presumably due to
density wave streaming motions, complicating the choice of PA.  We assume a PA
of $270\degr$ but test our results for sensitivity to this assumption.  The
kinematic center is determined to be at R.A. 02$^{\rm h}$42$^{\rm
m}$40.6$^{\rm s}$, Dec. -00$\degr$00$'$45$''$ (2000.0), and the systemic
velocity is $V_{sys,lsr}= 1130$ km s$^{-1}$ \citep{1995ApJ...450...90H}.  The
beam of the BIMA SONG data cube is 8.9''x5.5'' (PA 15.5$\degr$).  The
apertures are spaced by 6'' and are shown in Figure 10.

That NGC 1068 is dominated by molecular gas over the mapped region is
indicated by the global $M_{\rm H_2}/M_{\rm HI}$ value of 3.1 (using our value
of $X$), as well as a typical column density ratio of about 10 in the
pseudo-ring, as estimated from the BIMA SONG zeroth-moment map and from the
21-cm data of \citet{1997Ap&SS.248...23B}.

The resulting $<$$v$$>$--$<$$x$$>$ plot is shown in Figure 11.  The best fit
slope is $75 \pm 5$ km s$^{-1}$ kpc$^{-1}$, but the figure suggests that a
single pattern may not be a good description of the data, with apertures that
cross the region of the inner bar (8--11) indicating a somewhat steeper slope.
A fit to just these apertures yields a slope of $135 \pm 20$ km s$^{-1}$
kpc$^{-1}$.  A fit to the remaining apertures (that cross only the spiral)
yields $72 \pm 4$ km s$^{-1}$ kpc$^{-1}$.  As apertures 8--11 contain emission
from both the inner bar and the pseudo-ring and disk, if two patterns are
actually present with the outer pattern having the slower speed then the slope
for these apertures represents a lower limit to the inner pattern speed.
Thus, this result is at least consistent with the inner pattern speeds
determined by \citet{2000ApJ...533..850S} and \citet{1995ApJ...450...90H},
although the uncertainty is somewhat large.  However, the slope for the
remaining apertures is inconsistent with the outer pattern speed of
\citet{2000ApJ...533..850S}.

The $<$$v$$>$--$<$$x$$>$ plot for an assumed PA of 275$\degr$, shown in Figure
12a, shows much more scatter than the plot for the nominal PA, with best fit
slope to all apertures of $88 \pm 11$ km s$^{-1}$ kpc$^{-1}$.  For a PA of
265$\degr$, the best fit slope in the $<$$v$$>$--$<$$x$$>$ plot (Figure 12b)
is $55 \pm 5$ km s$^{-1}$ kpc$^{-1}$.  The scatter noticeably worsens for PAs
further from the nominal value, and we therefore conclude that the correct PA
is within this range.  We therefore allow that the PA may be uncertain by
5$\degr$.  As this is the dominant uncertainty in $\Omega_p$, we arrive at a
best fit value for all apertures of $75^{+13}_{-20}$ km s$^{-1}$ kpc$^{-1}$.
For apertures 8--11, taking into account the same uncertainty in PA, we find a
best fit inner pattern speed of $135 \pm 42$ km s$^{-1}$ kpc$^{-1}$, still
consistent with the two previous determinations but now with a large
uncertainty.  Similarly, using all apertures other than 8--11, we arrive at a
best fit outer pattern speed of $72^{+13}_{-18}$ km s$^{-1}$ kpc$^{-1}$.
Positive slope uncertainties correspond to positive PA uncertainties.

\begin{figure}
\epsscale{.80}
\plotone{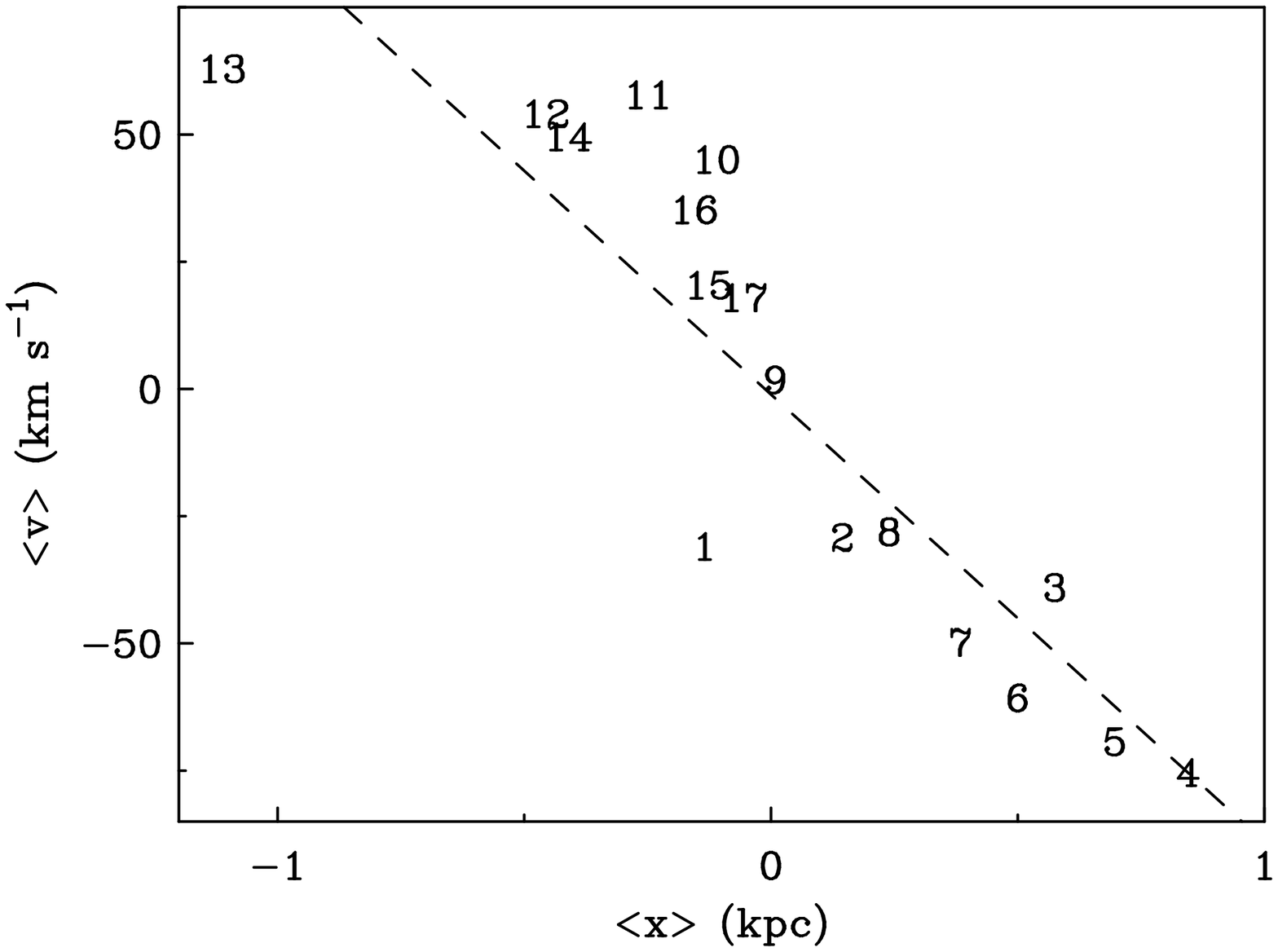}
\label{fig12}
\end{figure}

\begin{figure}
\epsscale{.80}
\plotone{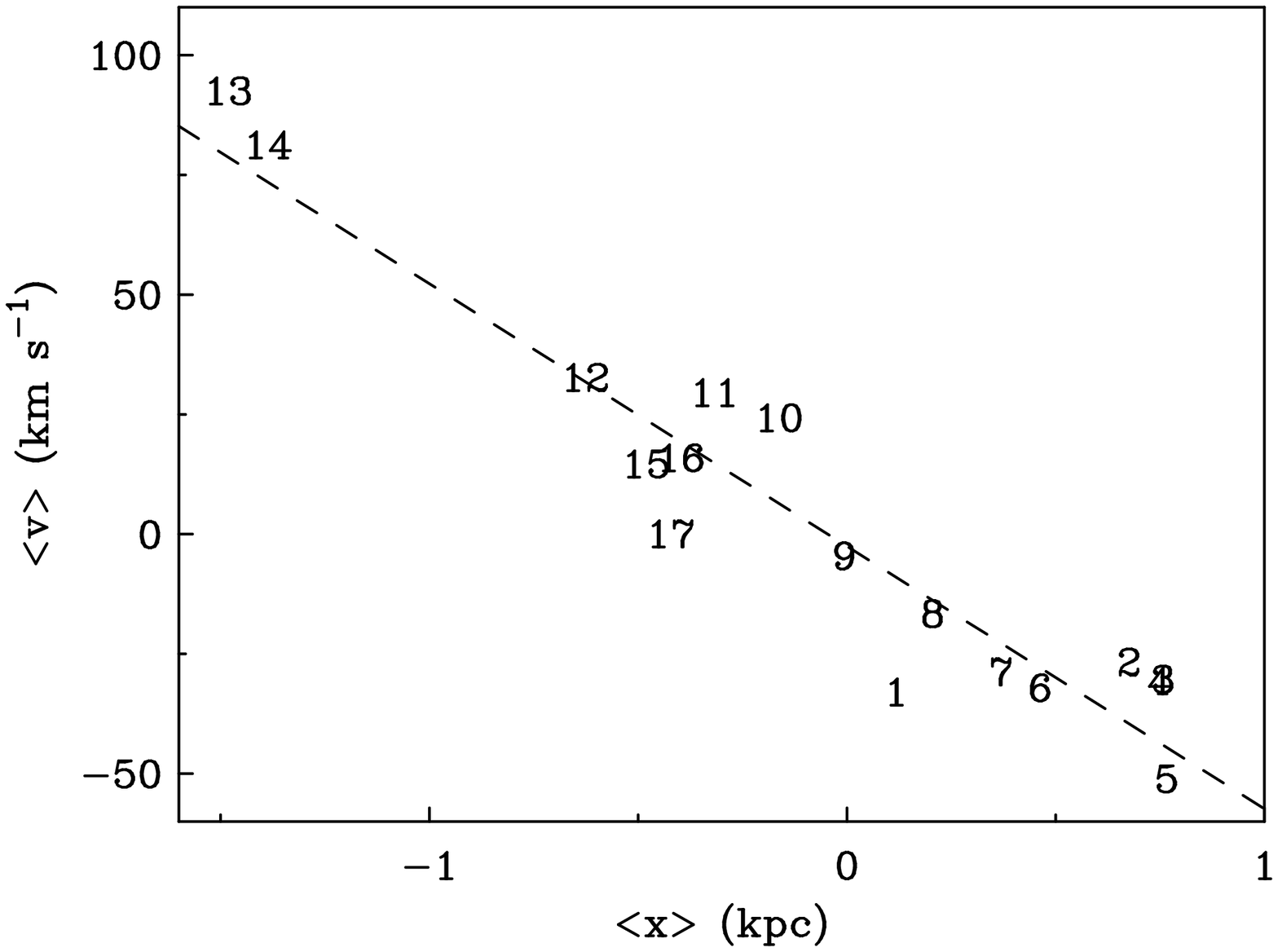}
\caption{Plot of $<$$v$$>$ vs. $<$$x$$>$ for the apertures shown in Figure 10,
but assuming a PA of a) 275 $\degr$ and b) 265$\degr$.
The dotted lines are the best-fit straight lines to all apertures.}
\end{figure}

Our results indicate that the spiral/pseudo-ring is not a response to the
inner bar pattern as found by \citet{1995ApJ...450...90H}; neither is it part
of an outer pattern as slow as the one argued for by
\citet{2000ApJ...533..850S}.  By inspecting the CO velocity field in Figure 3a
of \citet{2000ApJ...533..850S}, where the angular resolution is about three
times better than in the CO map of \citet{1995ApJ...450...90H}, the radial
streaming motions where the pseudo-ring crosses the minor axis on both sides
of the nucleus indicate an inward radial motion within the arms, which is as
expected for gas {\it inside} the CR of a pattern
[e.g. \citet{1990NYASA.596..130R}; the sense is also in agreement with the bar
model of \citet{2000ApJ...533..850S} where the pseudo-ring is part of the
outer pattern and located inside its CR; see below], not outside as argued for
by \citet{1995ApJ...450...90H}.  The streaming motions therefore support the
conclusion that the pseudo-ring is not part of the inner bar pattern.

In the interpretation of \citet{2000ApJ...533..850S}, the iILR and oILR of the
outer pattern roughly bound the spiral/pseudo-ring structure.  In their Figure
7, showing curves of $\Omega$, $\Omega - \kappa/2$ and $\Omega + \kappa/2$
vs. radius [or Figure 18 of \citet{1995ApJ...450...90H}, which is based on the
HI rotation curve of \citet{1997Ap&SS.248...23B}], pattern speeds of about 35
km s$^{-1}$ kpc$^{-1}$ or greater would indeed result in two ILRs which
bracket the pseudo-ring (although for their nominal choice of $20 \pm 10$ km
s$^{-1}$ kpc$^{-1}$ there would be no iILR and the oILR would be near $R=25''$
or greater).  In particular, a pattern with speed 72 km s$^{-1}$ kpc$^{-1}$ as
indicated by the TW method would have its iILR and oILR at 14'' and 19'',
respectively - a good match to the radial limits of the strong spiral
structure.  One should note, however, that there is some uncertainty,
estimated by \citet{2000ApJ...533..850S} to be $\pm 12$ km s$^{-1}$
kpc$^{-1}$, in the three curves in their Figure 7, which will affect the
location of the resonances.  It is also plausible that the CR of the inner
pattern, at $R=16-19''$ for pattern speeds of $140-160$ km s$^{-1}$
kpc$^{-1}$, coincides with one of the ILRs of the outer pattern, and this may
be an example of non-linear mode coupling
\citep{1987ApJ...318L..43T,1997A&A...322..442M,1999A&A...348..737R}, through
which energy and momentum is efficiently transferred from the inner to the
outer pattern, even if the patterns are relatively weak.  However, we note
that \citet{1999A&A...348..737R} conclude from their simulations that mode
coupling does not necessarily exist in all galaxies with multiple patterns.
We also note that, for a pattern of speed 72 km s$^{-1}$ kpc$^{-1}$,
corotation and the OLR of the outer pattern would be at roughly 50'' and 72'',
respectively [using Figure 18 of \citet{1995ApJ...450...90H}], with fairly
large uncertainties due to the shallow slopes of the $\Omega$ and $\Omega +
\kappa/2$ curves, such that the outer pattern, if it indeed has a radius of
120'', may extend well beyond its OLR.

The discrepancy between the outer pattern speeds from the TW method and
\citet{2000ApJ...533..850S} is unlikely to be due to the assumptions in the TW
method being violated.  The most likely cause for concern is the assumed value
of $X$ and thus the molecule-dominance of the region covered by the BIMA SONG
map.  However, the oxygen abundances of four HII regions with galactocentric
radii ranging from 20'' to 50'' are 12+log(O/H)$= 9.02 \pm 0.07$ \citep{
1987ApJ...319..662E}, close to the solar value of 8.81
\citep{1987soap.conf...89A}.  Based on the empirical calibration of $X$
vs. oxygen abundance from \citet{2002A&A...384...33B}, the choice of
the Galactic value would seem optimal, although there is considerable scatter.
Thus we conclude that NGC 1068 is indeed most likely molecule-dominated in the
region covered by the map.

\subsection{NGC 3627}

NGC 3627 (M66) is a member of the Leo Triplet (with NGC 3628 and NGC 3623),
and is classified as SAB(s)b in the RC3 catalog.  We adopt a distance of 11.1
Mpc, based on observations of Cepheid variables with the Hubble Space
Telescope \citep{1999ApJ...522..802S} .  A spectrum of the nucleus indicates
Liner/Seyfert 2 activity \citep*{1997ApJS..112..315H}.  HI observations
\citep*{1978AJ.....83..219R,1979ApJ...229...83H} indicate a tidal encounter
with NGC 3628, which has been modeled by \citet{1978IAUS...77..267T} and
\citet{1978AJ.....83..219R}.  The zeroth-moment CO map from the BIMA SONG data
at 12'' resolution (Figure 13) shows a two-arm barred spiral with an extension
of one arm to the south.  Here, and in near-IR images
\citep{2003AJ....125..525J}, the bar is long (about 2.6 kpc radius) and very
linear.  At full resolution, \citet{2002ApJ...574..126R} note in the BIMA SONG
data weak emission 20'' east and west of the bar which they interpret as
tracing an inner ring.  Emission from this structure is also seen in H$\alpha$
emission in the Fabry-Perot data of \citet{2003A&A...405...89C}.  From the
same data, these authors find overall kinematic parameters of $V_{sys,lsr}=
727 \pm 5$ km s$^{-1}$, $i=65\degr$, PA=$170 \pm 5\degr$ and a kinematic
center at R.A. 11$^{\rm h}$20$^{\rm m}$15$^{\rm s}$, Dec. 12$\degr$59$'$30$''$
(2000.0).  Unfortunately, the bar appears to be aligned very well with the
major axis, and therefore the TW method may not be applicable to it.  However,
if the strong spiral structure is generated by the bar potential with the same
pattern speed, then the TW method will indicate the speed of the bar/spiral.
The southern extended arm does not lie in the same plane as the rest of the
galaxy.  Beyond about $R=90''$, \citet{2003A&A...405...89C} find $i\approx
50\degr$ and PA$\approx 155\degr$ for this arm.

\begin{figure}
\epsscale{.80}
\plotone{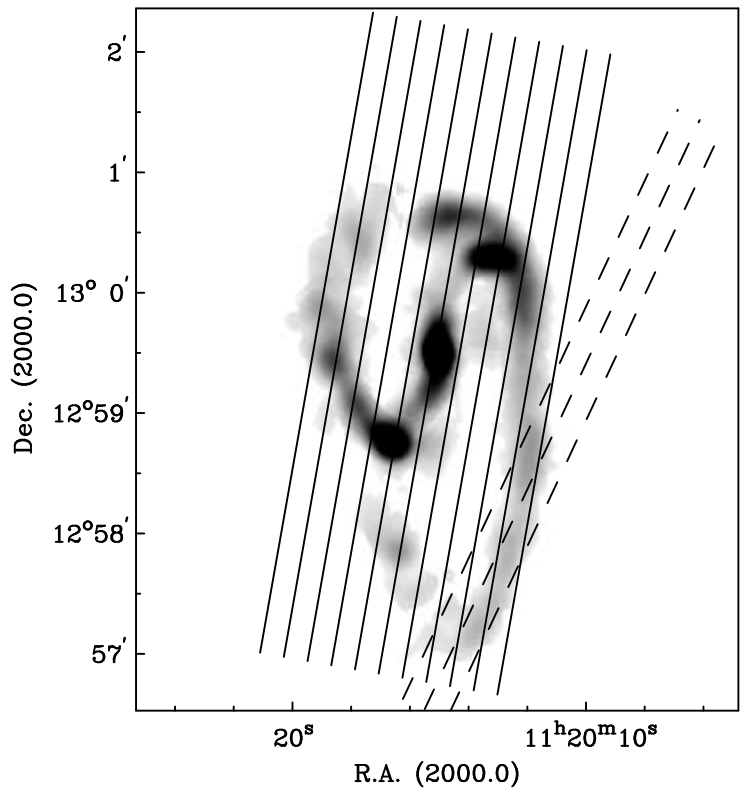}
\caption{Zeroth-moment map for NGC 3627.  Apertures used in the primary TW
calculation are shown as solid lines, using a PA of $170\degr$, with the
westernmost being Aperture 1.  The dashed lines show the three apertures used
for a separate application of the TW method for the southern extension of the
western arm using a PA of 155$\degr$.
\label{fig13}}
\end{figure}

\begin{figure}
\epsscale{.80}
\plotone{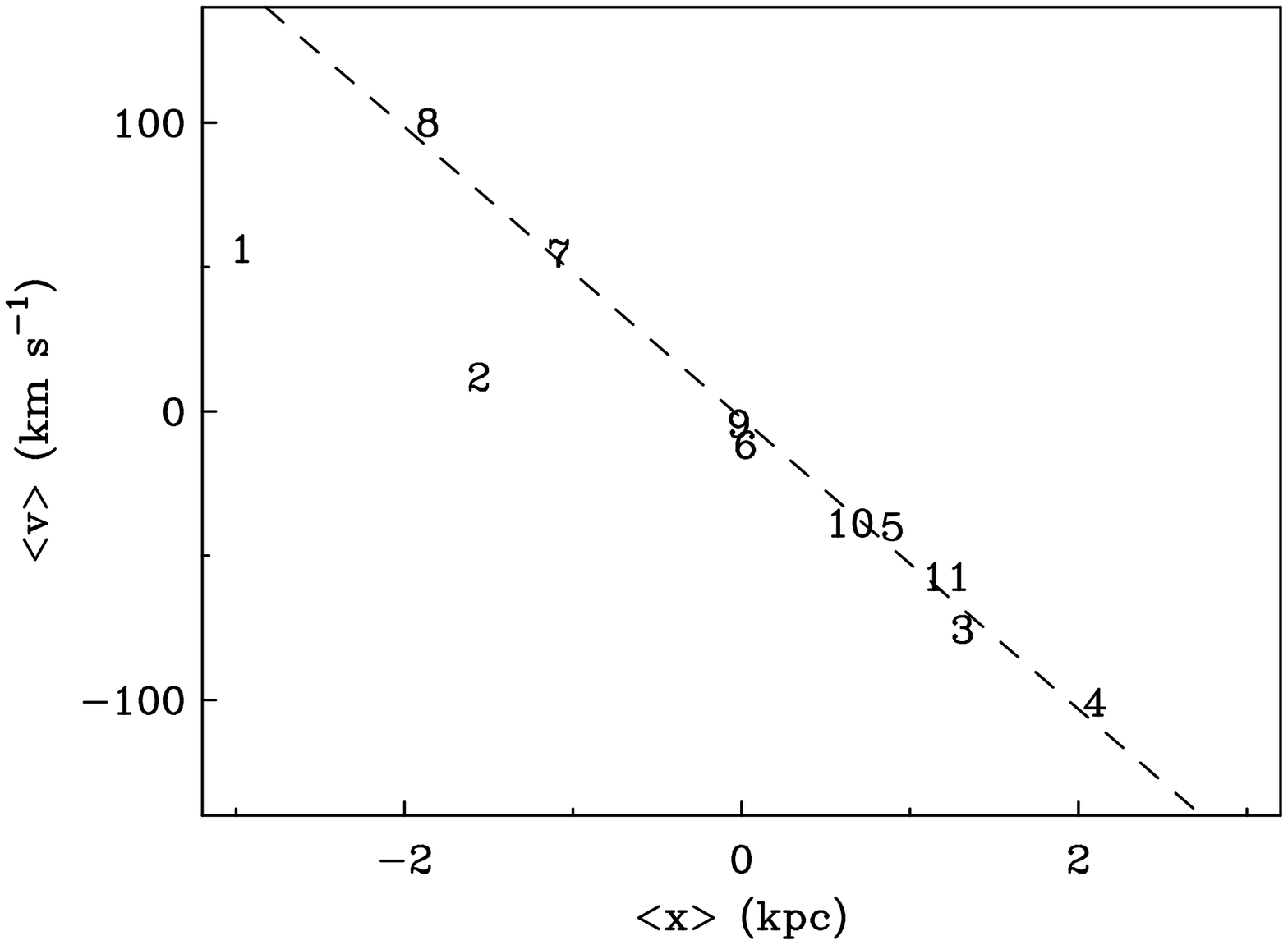}
\caption{Plot of $<$$v$$>$ vs. $<$$x$$>$ for the apertures shown in Figure 13
as solid lines.  The dotted line is the best-fit straight line to apertures
4--13.
\label{fig14}}
\end{figure}

Assuming that the CR of the bar is at 1--1.4 bar radii (49--69''),
\citet{2003A&A...405...89C} infer a bar pattern speed in the range $56-72$ km
s$^{-1}$ kpc$^{-1}$.  In support of this interpretation, they find evidence in
the residual velocity field for the signature of a change in sign of the
radial streaming motions induced by the pattern across the CR, as described by
\citet{1993ApJ...414..487C}, at $R=60-80''$.  The low end of this range of
$\Omega_P$ places the inner m=4 resonance at the location of the inner ring.
\citet{2002AJ....124.2581S} find a bar pattern speed of 55 km s$^{-1}$
kpc$^{-1}$ by similarly assuming that the bar radius is 80\% of the CR, but
using a CO rotation curve from \citet{2001AAS...199.5812D}.

NGC 3627 is strongly molecule-dominated.  The global molecular to atomic gas
mass ratio within the optical disk is estimated by \citet{1996A&A...306..721R}
to be about 7 (using our value of $X$).  The radial profiles of
\citet*{1993ApJ...418..100Z} indicate that the molecular-to-atomic gas
column density ratio never drops below 7 within the detectable limit of
emission in Figure 13.

The TW method was applied to moment maps made from the BIMA SONG data cube
smoothed to 12'' resolution.  The 12''-spaced apertures used are shown for
PA=170$\degr$ in Figure 13 as the solid lines, and the $<$$v$$>$--$<$$x$$>$
plot is shown in Figure 14.  Apertures 1--3 are dominated by emission from the
southern extension of the western arm, where $i$ and PA are both determined to
be 15$\degr$ smaller than for the rest of the disk, as discussed above.  They
obviously do not fit the relation as defined by the remaining apertures, and
we consider them separately.  The best fit slope to apertures 4--11 is $50 \pm
2$ km s$^{-1}$ kpc$^{-1}$.  Allowing for the uncertainty in the PA quoted by
\citet{2003A&A...405...89C}, the best fit slope is $50^{+3}_{-8}$ km s$^{-1}$
kpc$^{-1}$, with positive uncertainties corresponding to positive PA
uncertainties.  This value is in reasonable agreement with the previous
determinations of the bar's pattern speed quoted above, and suggests that the
spiral and bar may indeed form a pattern with a single speed.  If so, then
only the bright peaks at the very beginnings of the arms are inside the CR,
which, using the rotation curve of \citet{2003A&A...405...89C}, would fall at
$R=80''$.  Bright, narrow two-armed structure extends beyond this radius to
about $R=120''$ east and west of the nucleus in the plane of the galaxy
(Figure 13).  There is a hint of a similar extent in the near-IR
\citep{2003AJ....125..525J}, although the emission is faint.  Such a situation
is found generally for galaxies with large bars by
\citet{1995ApJ...445..591E}.

For the southern extension of the western arm, using the three apertures for
PA=155$\degr$, shown as dashed lines in Figure 13, a best fit slope of $23 \pm
4$ km s$^{-1}$ kpc$^{-1}$ is found.  Allowing for a $\pm 5\degr$ uncertainty
in the PA does not increase the uncertainty in the slope significantly.
However, the result is more uncertain than the formal error bar would indicate
as it is only based on three apertures, at least two of which probably contain
emission from gas participating in both patterns, if the outer pattern begins
at the radius where $i$ and PA begin to change.  We therefore conclude only
tentatively that this arm segment may represent a distinct pattern.

\subsection{NGC 4321}

NGC 4321 (M100) is a well-studied grand-design spiral galaxy in the Virgo
Cluster, carrying a classification of SAB(s)bc in the RC3 catalog.  We use the
Cepheid-based distance of 16.1 Mpc \citep{1996ApJ...464..568F}.  The central
region includes a bar of length $\sim 15''$ and PA $153 \pm 3\degr$
(e.g. \citealt{1995AJ....110.2075S,1995A&A...296...45S,
1995ApJ...454..623K,1995ApJ...443L..73K}; \citealt*{1998ApJ...494..236W};
\citealt{2000ApJ...528..219K}).  Using hydrodynamical simulations and a
potential derived from an analytical bar model, \citet{1998ApJ...494..236W}
conclude that a bar with a pattern speed of about 70 km s$^{-1}$ kpc$^{-1}$
produces gas morphology and kinematics in best agreement with CO data.  For
the disk, \citet{1995A&A...296...45S} used two methods to estimate the pattern
speed.  The first is the method of \citet[see above]{1993ApJ...414..487C},
through which a value of about 29 km s$^{-1}$ kpc$^{-1}$ (scaled to the
inclination of 27$\degr$ that we adopt below) was found.  The second involves
numerical hydrodynamic simulations of the response of a cloud population to a
potential derived from an image of the galaxy in red light.  This method
indicated a pattern speed of about 25 km s$^{-1}$ kpc$^{-1}$, which must also
be scaled to account for the different assumed inclinations (they assumed
$i=32\degr$), although this scaling is presumably model-dependent.
\citet{1989ApJ...343..602E} find that a pattern with a speed of about 27 km
s$^{-1}$ kpc$^{-1}$ (again scaled to our adopted inclination) for the disk has
its ILR and OLR roughly bounding the extent of the spiral structure, with gaps
in the I-band light at the 4:1 resonance and the CR.
\citet{1995AJ....109.2444R} estimates a value of $\lesssim 35$ km s$^{-1}$
kpc$^{-1}$ by identifying the CR with the location where the southern CO arm
in his BIMA map crosses the major axis, where tangential streaming motions all
but vanish.  \citet{2002AJ....124.2581S} identify a large bar of radius 54''
in a $K$-band image (see also \citealt{2003ApJ...582..190D}) associated with
the inner molecular spiral structure.  Using the same assumptions as for NGC
3627, they derive a pattern speed of 40 km s$^{-1}$ kpc$^{-1}$ (again scaled
to our assumed inclination).  If the bulk of the CO emission morphology
outside the inner bar is due to this stellar bar potential, then
\citet{2002AJ....124.2581S} should be measuring the same pattern as in the
other studies.

\begin{figure}
\epsscale{.80}
\plotone{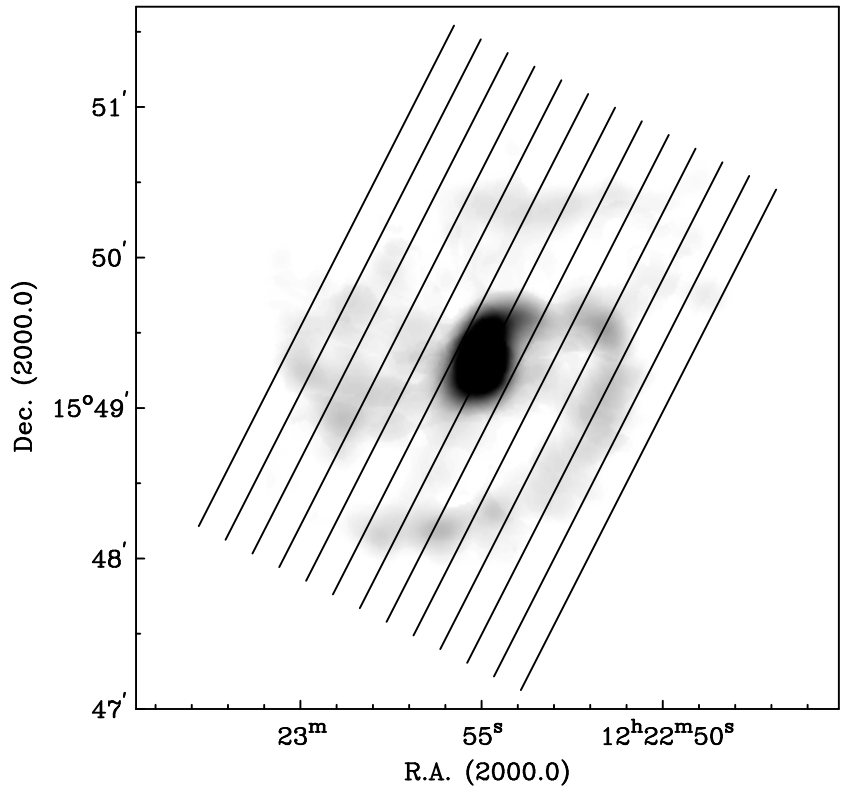}
\caption{Zeroth-moment map for NGC 4321.  Apertures used in the TW
calculation are shown, with the westernmost being Aperture 1.
\label{fig15}}
\end{figure}

\begin{figure}
\epsscale{.80}
\plotone{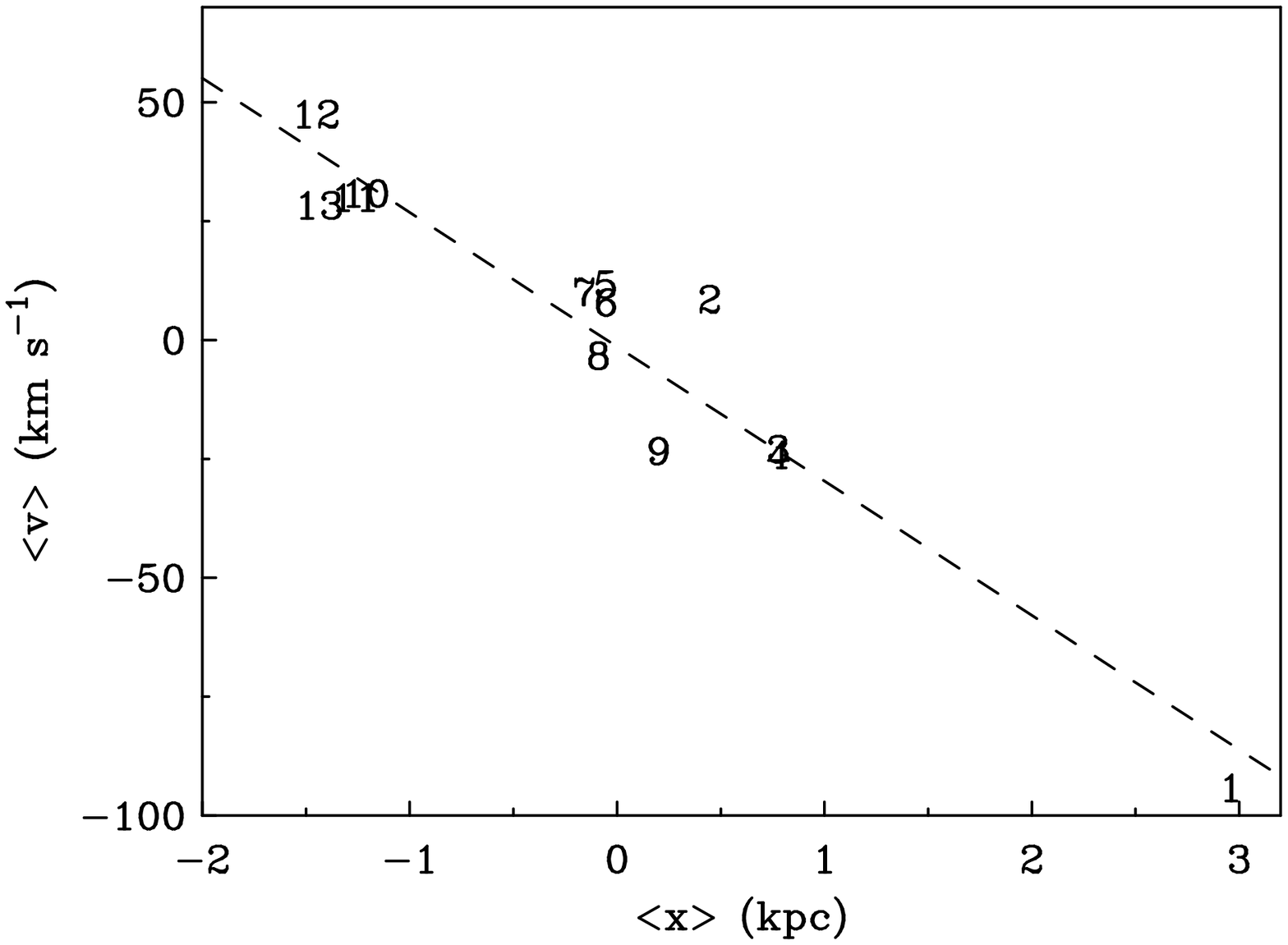}
\caption{Plot of $<$$v$$>$ vs. $<$$x$$>$ for the apertures shown in Figure 15.
The dotted line is the best-fit straight line to all apertures.
\label{fig16}}
\end{figure}

We use the kinematic center of R.A. 12$^{\rm h}$22$^{\rm m}$55.0$^{\rm s}$,
Dec. 15$\degr$49$'$20$''$ (2000.0), PA of 153$\degr$, and $V_{sys,lsr}= 1575$
km s$^{-1}$ from the CO study of \citet{1997A&A...325..769S}, and $i=27\degr$
from the analysis of VLA 21-cm data by \citet{1993ApJ...416..563K}.  The
kinematic center and systemic velocity agree well with values derived from
interferometric CO data by \citet{1995AJ....110.2075S}.  We found uncertainty
in the results was decreased significantly after smoothing the data cube to
12'' resolution.  The 12''-spaced apertures used are shown in Figure 15.  Note
that the nominal major axis PA is equal to that of the inner bar, and thus it
should have no signature in the TW method.

At the detection limit of CO in the BIMA SONG map, at about $R=80''$, we
estimate the ratio of molecular to atomic column density to be about 1.2--1.7
(using our value of $X$), the range largely determined by whether the CO data
used derive from the BIMA SONG zeroth-moment map, or the single-dish data of
\citet{1996A&A...308...27K}, \citet{1988ApJS...66..261K}, or
\citet{1997A&A...325..769S}.  The HI data are from
\citet{1993ApJ...416..563K}.  Within this radius, the CO intensity rises while
the 21-cm intensity stays flat, eventually falling in the central regions.
Hence, we assume this galaxy is molecule-dominated over the region of
detectable CO in the BIMA SONG map.

In Figure 16, we show the $<$$v$$>$--$<$$x$$>$ plot.  The best fit slope is
$28 \pm 3$ km s$^{-1}$ kpc$^{-1}$.  Aperture 1, which has somewhat extreme
values of $<$$v$$>$ and $<$$x$$>$ relative to Apertures 2--10 may
significantly bias the fit.  However, without this aperture, the slope becomes
$25 \pm 4$ km s$^{-1}$ kpc$^{-1}$, equal to the above value within the
uncertainties.  It is very difficult to put constraints on possible errors in
the PA.  An analysis of the rotation curve using the GIPSY program ROTCUR on
the first-moment map, fixing the systemic velocity, inclination and kinematic
center at their above values, indicates a very constant PA of $153-155\degr$
at $R=50-90''$.  Inside this radial range, the PA is generally larger but
varies greatly, between 144$\degr$ and 171$\degr$, due to very strong
streaming motions which are obvious in the first-moment map of, e.g.,
\citet{1997A&A...325..769S}.  Beyond $R=90''$ there are few data points.  If
we allow for an uncertainty of $3\degr$ in the PA, the best fit slope becomes
$28^{+4}_{-5}$ km s$^{-1}$ kpc$^{-1}$, where the error bars incorporate the PA
uncertainty and the scatter in the $<$$v$$>$--$<$$x$$>$ plot for the nominal
PA.  Positive slope uncertainties correspond to positive PA uncertainties.
This value agrees within the error bars with the above pattern speed
determinations for the disk by \citet{1995A&A...296...45S} and
\citet{1989ApJ...343..602E}, using three independent methods, and is somewhat
lower than the upper limit found by \citet{1995AJ....109.2444R} and the
large-scale bar pattern speed inferred by \citet{2002AJ....124.2581S}.  We
therefore conclude that the pattern speed for the disk of NGC 4321 is fairly
robustly determined.

\subsection{NGC 4414}

NGC 4414 is a flocculent spiral classified as SA(rs)c? in the RC3 catalog.
\citet{1997ApJ...490..682T} analyze a $K'$-band image, making an attempt to
correct it for contamination by supergiants, and find that despite the
flocculent optical appearance there are smooth, kpc-scale coherent features in
the old stellar disk.  Most noticeable is a patchy ring structure of radius
20''.  However, it is unclear what large-scale dynamical influence might be
responsible for these features.  Hence it is not clear whether large-scale
density waves exist in this galaxy, let alone whether there is only one
pattern or more.  The first-moment map (their Figure 9) indicates a very
regular velocity field, lacking the characteristic wiggles in the iso-velocity
contours that would suggest density wave streaming motions.  Furthermore, they
find from a BIMA (combined with single-dish data) map at about 3'' resolution
that, although the flocculent molecular structure shows some correlation with
the coherent $K'$-band structure (especially the ring evident in Figure 17),
there are many molecular peaks which do not follow this relation, suggesting a
significant role for stochastic effects.  The TW method can of course be
attempted, but the result may not be identifiable as a pattern speed.

We adopt a Cepheid-based distance of 19.1 Mpc \citep{1998ApJ...505..207T}.
The CO and HI velocity fields were studied by \citet{1997ApJ...490..682T}.
Their CO rotation curve analysis indicates $i=60\degr$, PA=160$\degr$, and
$V_{sys,lsr}= 726$ km s$^{-1}$.  We also adopt their central position of
R.A. 12$^{\rm h}$26$^{\rm m}$27.1$^{\rm s}$, Dec. 31$\degr$13$'$24$''$
(2000.0).  The beam for the BIMA SONG data is 6.4''x5.0'' (PA 5$\degr$).
Figure 17 shows the zeroth-moment map and the apertures used.  The apertures
are spaced by 6''.

Figure 8 of \citet{1997ApJ...490..682T} demonstrates that the ISM of NGC 4414
is molecule-dominated.  The ratio of molecular to atomic surface density (with
our value of $X$) is everywhere $>4$ over the region where emission is
detected in the BIMA SONG map.

\begin{figure}
\epsscale{0.9}
\plotone{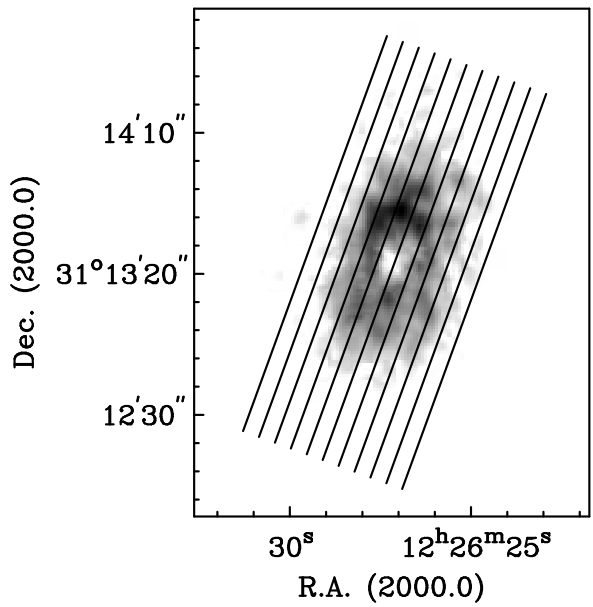}
\caption{Zeroth-moment map for NGC 4414.  Apertures used in the TW
calculation are shown, with the westernmost being Aperture 1.
\label{fig17}}
\end{figure}

\begin{figure}
%\epsscale{.80}
\plotone{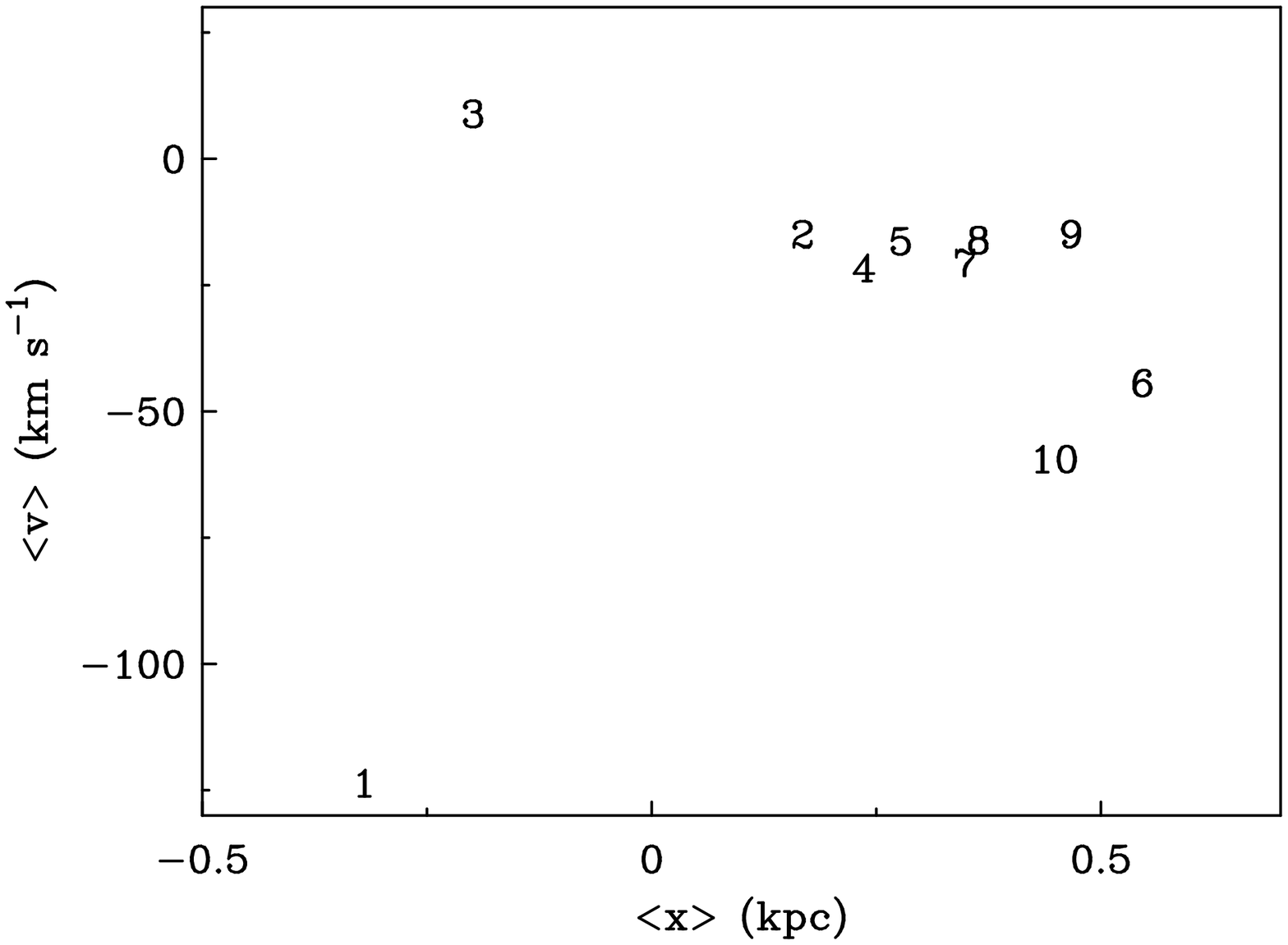}
\caption{Plot of $<$$v$$>$ vs. $<$$x$$>$ for the apertures shown in Figure 17.
\label{fig18}}
\end{figure}

Figure 18 shows the $<$$v$$>$--$<$$x$$>$ plot.  There is no well defined slope
and thus no indication of a simple pattern in this galaxy.  Even eliminating
the discrepant Aperture 1, a fit to the remaining apertures yields a very
uncertain slope of $66 \pm 22$ km s$^{-1}$ kpc$^{-1}$.  The points are almost
all found in positive-$<$$x$$>$, negative-$<$$v$$>$ quadrant because the
patchy ring is brighter on the approaching side of the galaxy for almost all
apertures.  Better correlations in the $<$$v$$>$--$<$$x$$>$ plot start to
appear for PA's of only $\pm 5\degr$ away from the nominal value, but the
ROTCUR analysis of \citet{1997ApJ...490..682T} (their Figure 11) indicates
that the PA is well determined to within about $\pm 2\degr$ or so, due to the
very regular velocity field.  The correlations are therefore certainly due to
the PA error, as found for the simulations of the tilted, axisymmetric clumpy
disk (\S 3.1), which this galaxy's CO distribution resembles.  Given the lack
of evidence for a pattern in this galaxy, we cannot conclude that we have
measured the speed of any pattern.

\subsection{NGC 4736}

NGC 4736 (M94) is classified as (R)SA(r)ab in the RC3 catalog and is also
classified as a LINER \citep{1980A&A....87..152H}.  We adopt the distance of
4.2 Mpc used by \citet{2000ApJ...540..771W}.  The galaxy features a bar of
radius $\sim 20''$, seen in the near-IR \citep*{1995A&A...301..359M} and CO
(Figure 19; \citealt{2000ApJ...540..771W,1999ApJS..124..403S}), and a ring of
star formation of radius $\sim 45''$ visible in the H$\alpha$ image of
\citet{1989ApJS...71..433P}.  While quite ringlike in H$\alpha$ emission, the
corresponding structure in CO is a pair of tightly wrapped spiral arms (see
Figure 19 of \citealt{2000ApJ...540..771W}). Optical images also reveal a
faint outer stellar ring of radius $\sim 5'$.  \citet*{1991A&A...251...32G}
find that the disk is nonaxisymmetric, and suggest that the spiral/ring and
the outer ring may be located at the ILR and OLR of an oval potential,
deriving a pattern speed of 45 km s$^{-1}$ kpc$^{-1}$.
\citet{1995A&A...301..359M} suggest, based on their bar models and those of
\citet{1981ApJ...247...77S}, that the inner spiral/ring represents gas
transported outward by the bar to near its OLR (thus its OLR roughly coincides
with the ILR of the outer pattern), and thus find a bar pattern speed of about
460 km s$^{-1}$ kpc$^{-1}$.  \citet{2000ApJ...540..771W} analyze CO (from the
BIMA SONG data) and HI rotation curves and conclude that, if the gas
spiral/ring is at the OLR of the bar, its pattern speed must also be high,
around 400 km s$^{-1}$ kpc$^{-1}$.  CR is at about 1.3 bar radii, or 26'', for
this speed.  They also find about the same pattern speed for the oval
distortion as found by \citet{1991A&A...251...32G}, under the same assumptions
regarding the ILR and OLR, but the uncertainties are large due to the poorly
determined HI rotation curve in the outer disk.  Furthermore,
\citet{2000ApJ...540..771W} note that the deprojected bar and oval are neither
aligned nor orthogonal, and thus cannot be sustained by a single pattern as
shown by \citet{1988MNRAS.233..337L}.

We take kinematic parameters from the CO rotation curve analysis of
\citet{2000ApJ...540..771W}: $V_{sys,lsr}=315$ km s$^{-1}$, R.A. 12$^{\rm
h}$50$^{\rm m}$53.1$^{\rm s}$, Dec. 41$\degr$07$'$14$''$ (2000.0),
PA=$295\degr$ (although it rises from $290\degr$ to $300\degr$ between
$R=25''$ and $R=60''$).  They adopted the photometric inclination of 35$\degr$
from \citet{1995A&A...301..359M} which we will also use.  The beam for the
BIMA SONG data is 6.9''x5.0'' (PA 63$\degr$).

The radial profiles of \citet{2000ApJ...540..771W} show that the galaxy is
molecule-dominated in the region detected in CO in the BIMA SONG survey.
Almost all of the emission is within 1' from the center, where the ratio of
molecular to atomic hydrogen column density is 2 or greater (using our value
of X).

The apertures used are shown in Figure 19, and the resulting
$<$$v$$>$--$<$$x$$>$ plot is shown in Figure 20.  A clear correlation is
present.  The best fit slope to all apertures is $185 \pm 16$ km s$^{-1}$
kpc$^{-1}$.  Apertures 9--13 cross the central bar and may be affected by more
than one pattern.  In Figure 20, it is clear that a line fit to these
apertures would have a higher slope.  For these apertures alone, the best fit
slope is $356 \pm 139$ km s$^{-1}$ kpc$^{-1}$.  For the remaining apertures,
the best fit slope is $166 \pm 13$ km s$^{-1}$ kpc$^{-1}$.

\begin{figure}
\epsscale{.80}
\plotone{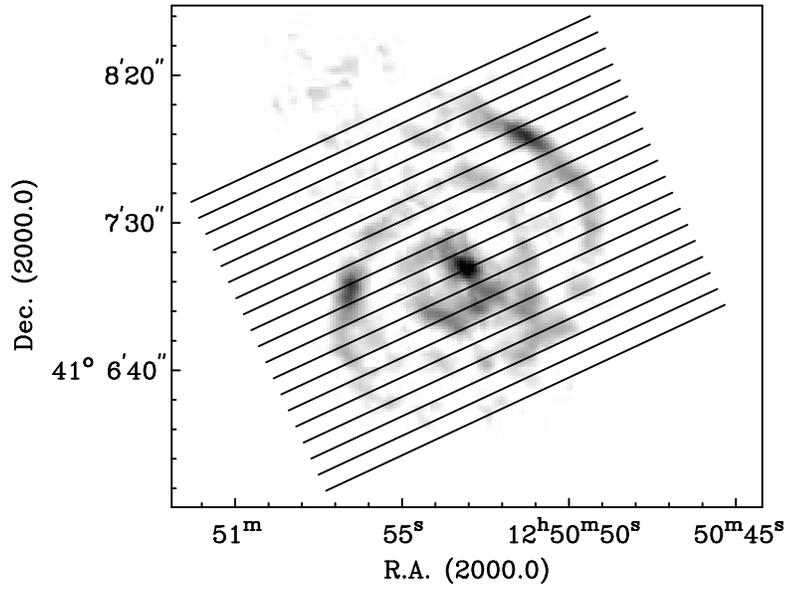}
\caption{Zeroth-moment map for NGC 4736.  Apertures used in the TW
calculation are shown, with the northernmost being Aperture 1.
\label{fig19}}
\end{figure}

\begin{figure}
\epsscale{.80}
\plotone{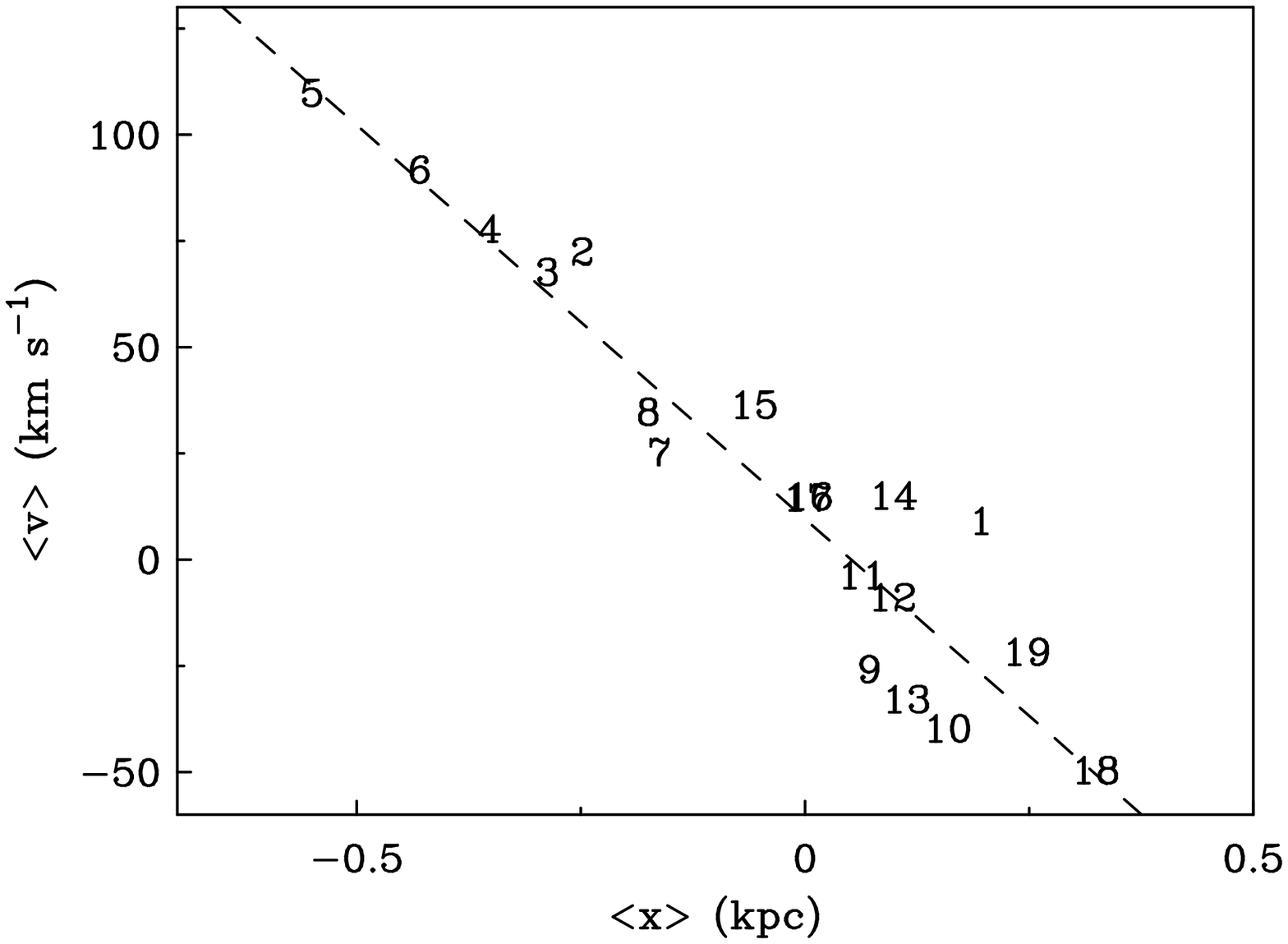}
\caption{Plot of $<$$v$$>$ vs. $<$$x$$>$ for the apertures shown in Figure 19.
The dotted line is the best-fit straight line to all apertures.
\label{fig20}}
\end{figure}

We assess the effects of PA errors by considering the above 10$\degr$ range of
PA found for the radial range of detected CO emission by
\citet{2000ApJ...540..771W}.  For all apertures, the best fit slope becomes
$185 \pm 34$ km s$^{-1}$ kpc$^{-1}$, with larger values corresponding to
smaller choices of PA.  For apertures 9--13 alone, the uncertainties due to
this range of possible PA is insignificant compared to the formal uncertainty
above.  For all apertures other than 9--13, the best fit slope becomes
$166^{+26}_{-32}$ km s$^{-1}$ kpc$^{-1}$.

The value for the central apertures is at least consistent with the fast bar
argued for by \citet{2000ApJ...540..771W}, although the uncertainties are
large and emission in these apertures may arise from gas participating in more
than one pattern.  The value for the disk outside the central bar is much
larger than the pattern speed of 45 km s$^{-1}$ kpc$^{-1}$ argued for by
\citet{1991A&A...251...32G}.  Using the resonance diagram (Figure 9) of
\citet{2000ApJ...540..771W}, this larger pattern speed would place CR just
outside the gas spiral/ring, at about $R=50-70''$.  The OLR would be at about
$R=70-120''$.  There may be one or two ILRs, at around $R=20''$, but these
placements are uncertain due to the large error bars in the determination of
the $\Omega - \kappa/2$ curve.

Nevertheless, can a sensible explanation of the resonance locations and
associated structures be found for this larger pattern speed?  Within the
large uncertainties, the ILR(s) of the outer pattern may coincide with the CR
of the inner bar, thus making this galaxy another potential example of mode
coupling.  Given these uncertainties, coupling involving coincidence of the
inner 4:1 resonance of the outer pattern with the CR of the inner one, which
is found by \citet{1999A&A...348..737R} to be more common than CR-ILR
coupling, is also possible.

%R+SALO 2000 FIND CASES WHERE OLR OF BAR HAS A RING, EVEN THOUGH A SLOWER
%SPIRAL IS PRESENT.  SO COULD SPIRAL/RING BE A FEATURE OF BAR PATTERN ONLY?

The CR of the outer pattern would be just outside the bright CO emission and
star formation.  \citet{1989ApJ...343..602E} argued for this arrangement for
M81 and M100, and indeed their derived pattern speed for M100 agrees with the
two estimates of \citet{1995A&A...296...45S} and our analysis above.  The
outer stellar ring at $R\approx 5'$ may not be associated with any resonance
of these patterns and could potentially be due to a third pattern.  However,
it may have a more natural explanation if it is indeed at the OLR of a 45 km
s$^{-1}$ kpc$^{-1}$ outer pattern, as the ratio of its radius to the
spiral/ring radius is close to the ratio of the OLR to ILR radius of 5.8 for a
single pattern with a flat rotation curve \citep{2000ApJ...540..771W}.
Nevertheless, the assumptions for the TW method are quite well satisfied for
this galaxy (especially for the apertures that do not cross the bar).  As for
NGC 1068, the most likely concern is that $X$ may be lower for this galaxy
than assumed, calling into doubt the molecular-dominance of the region covered
by the BIMA SONG map.  However, the oxygen abundances of nine HII regions with
galactocentric radii ranging from 35'' to 135'' are in the range 12+log(O/H)$=
8.86-8.98$ \citep{1993ApJ...411..137O}, again close to the solar value of 8.81
\citep{1987soap.conf...89A}, suggesting that the choice of the Galactic value
of $X$ is reasonable.  Hence, we conclude that the ISM of the region we
analyze is indeed most likely dominated by molecular gas, and thus our
application of the TW method should be well justified.

\subsection{NGC 4826}

NGC 4826 (M64) is classified as (R)SA(rs) in the RC3 catalog.  It is also well
known as the Evil Eye galaxy as a result of its dust morphology, and features
an outer disk of gas which rotates in the opposite sense to the stars and gas
in the inner disk
(\citealt*{1992Natur.360..442B,1993PASJ...45L..47V,1993A&A...279L..41C};
\citealt{1994ApJ...420..558B,1995ApJ...438..155R}).
\citet{1994ApJ...420..558B} find that the inner disk shows well ordered
rotation within 1' (1.1 kpc at their assumed distance of 3.8 Mpc, which we
also assume here) of the nucleus, outside of which is a region of disturbed
kinematics and, beyond a radius of 200'', the counterrotating gas disk seen in
21-cm emission.  These authors conclude that the counterrotating HI gas was
accreted during an interaction or in a continuous process.
\citet{1994ApJ...420..558B} find that the intermediate region shows a
signature of infalling HI gas, although \citet{1995ApJ...438..155R} find that
the ionized gas kinematics in this region are more complex than a simple
infall model would predict.  Our zeroth-moment map from the BIMA SONG data
(Figure 21) reveals CO emission out to a radius of about 50'', and thus all
the molecular gas detected is in the inner disk.
\citet{2003A&A...407..485G} observe CO 1--0 and 2--1 emission with the IRAM
Plateau de Bure interferometer and find evidence for a prominent $m=1$ mode in
the gas in the inner 1 kpc through the observation of two molecular arms on
opposite sides of the galaxy occupying distinct radial ranges (11--16'' and
16--33'', respectively).  The inner arm shows evidence of radial outflow, while
the outer arm shows a radial inflow signature, and thus the authors suggest
that the inner arm is located outside the CR of a fast mode ($\sim 1500$ km
s$^{-1}$ kpc$^{-1}$; placing CR near 100 pc, well inside the inner arm) while
the outer arm is part of a stationary mode, which is found in simulations of
galaxies with counter-rotation \citep{2000A&A...363..869G}.  The F450W
Hubble Space Telescope image of \citet{2003A&A...407..485G} suggests a
rather flocculent spiral in the central 1 kpc.

We adopt a PA of $112\degr$, inclination of 60$\degr$, dynamical center of
R.A. 12$^{\rm h}$56$^{\rm m}$43.7$^{\rm s}$, Dec. 21$\degr$40$'$59$''$
(2000.0), and $V_{sys,lsr}=416$ km s$^{-1}$ from the CO kinematic study of
\citet{2003A&A...407..485G}.  The beam for the BIMA SONG data is 7.5''x5.2''
(PA 3$\degr$).

\begin{figure}
\plotone{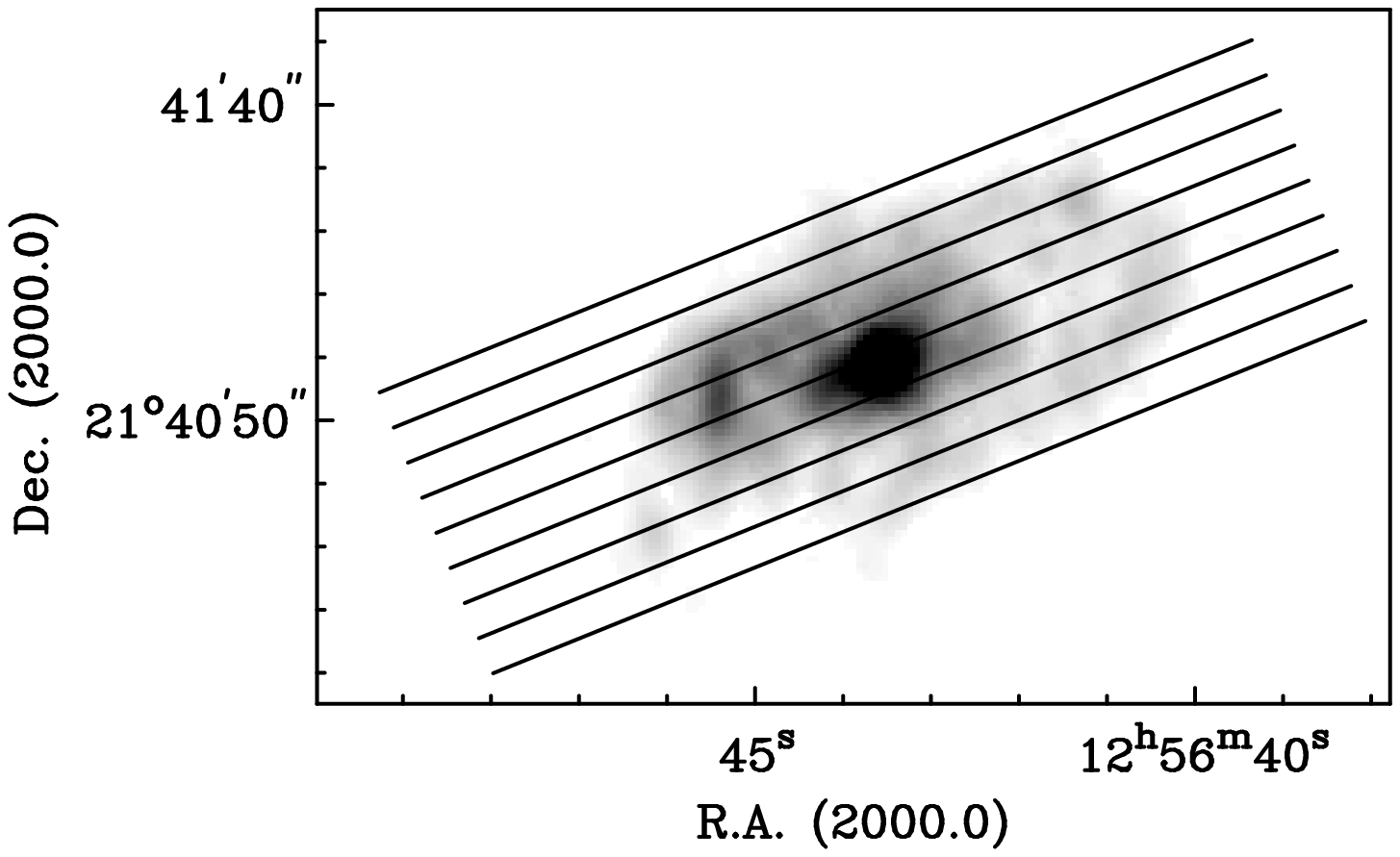}
\caption{Zeroth-moment map for NGC 4826.  Apertures used in the TW
calculation are shown, with the northernmost being Aperture 1.
\label{fig21}}
\end{figure}

\begin{figure}
\epsscale{.80}
\plotone{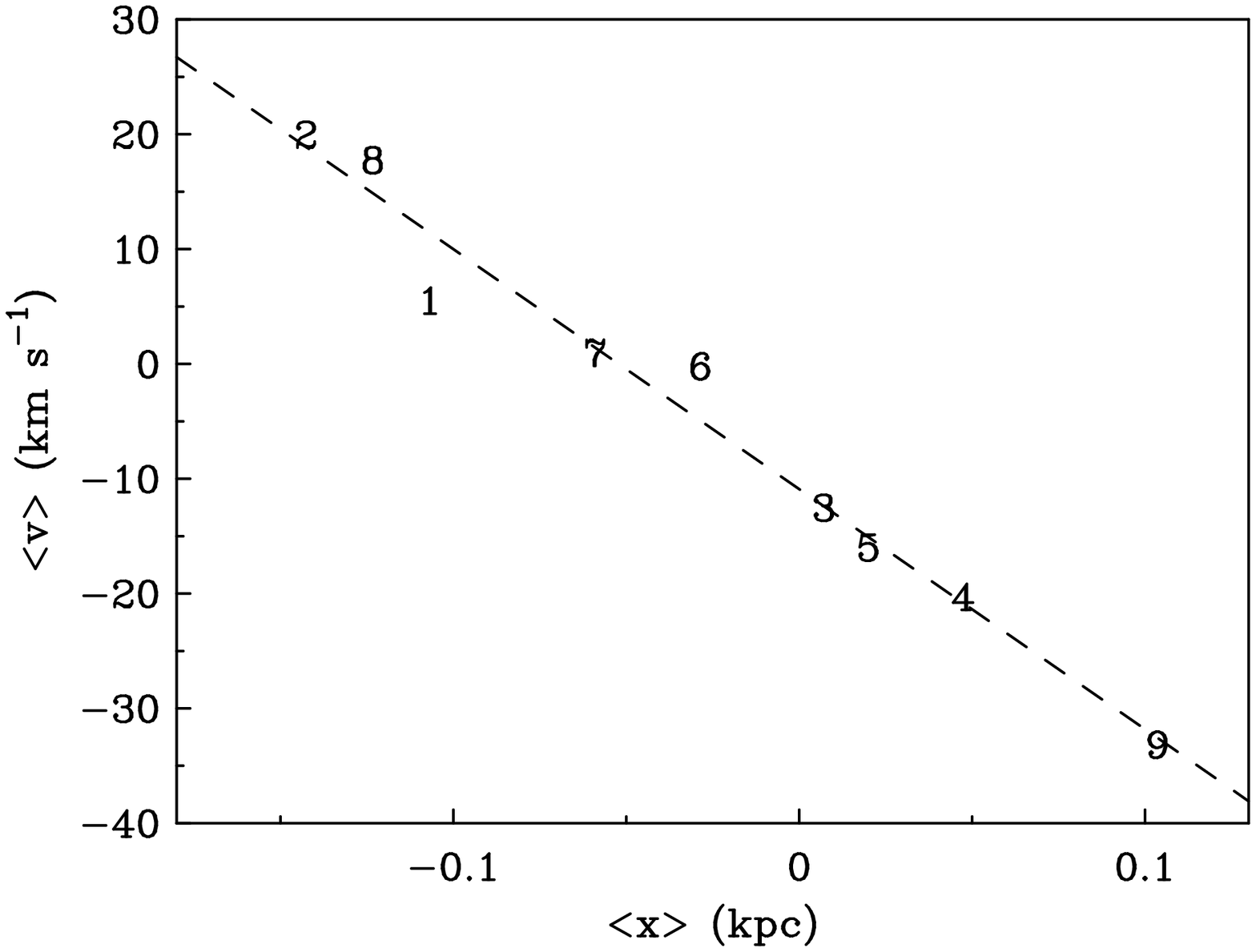}
\caption{Plot of $<$$v$$>$ vs. $<$$x$$>$ for the apertures shown in Figure 21.
The dotted line is the best-fit straight line to all apertures.
\label{fig22}}
\end{figure}

In the normally rotating inner disk, \citet{1993A&A...279L..41C} measured a
molecular mass of $1.8\times 10^8$ M$\sun$ (scaled to our value of $X$) with
the IRAM 30-m telescope, while the HI mass in the same region is $1.1 \times
10^7$ M$\sun$ \citep{1994ApJ...420..558B}, so that the region of the ISM over
which CO is detected in the BIMA SONG map (within $R=50''$) is clearly
molecule-dominated.

The apertures used are shown in Figure 21, overlaid on a zeroth-moment map of
the BIMA SONG data cube.  The resulting $<$$v$$>$--$<$$x$$>$ plot is shown in
Figure 22.  There is a well-defined slope whose best fit value is $209 \pm 13$
km s$^{-1}$ kpc$^{-1}$.  However, many pieces of evidence suggest that there
is no pattern here.  First, there is no clear sequence of aperture numbers in
Figure 22, suggesting that random clumpiness may dominate the integrals used
to calculate $<$$v$$>$ and $<$$x$$>$.  The values of $<$$v$$>$ and $<$$x$$>$
are also small, with absolute values of $<$$x$$>$ reaching no more than 8''
even for the apertures furthest from the major axis.  Second, for our
simulated clumpy, axisymmetric disk, we found that the slope in the
$<$$v$$>$--$<$$x$$>$ plot reflected the slope of the rotation curve.  We have
used ROTCUR to determine a rotation curve from a first-moment map.  We exclude
points within $20\degr$ of the minor axis, and fix the kinematic center,
systemic velocity, PA and inclination at their above values.  We also assume
no expansion.  We fit and plot approaching and receding sides separately to
allow the general velocity gradient to be seen.  The resulting rotation curve
is shown in Figure 23.  Also plotted are lines representing the best fit
linear slope of the rotation curve of $244 \pm 21$ km s$^{-1}$ kpc$^{-1}$, and
the best fit slope for the data in Figure 22.  These slopes agree well.
Therefore, the best fit slope may simply be reflecting the gradient in the
rotation curve.  Third, there is no evidence for any strong bar or spiral
structure in near-IR images of NGC 4826 \citep{2001A&A...368...16M}.
Therefore, despite the evidence for radial motions presented by
\citet{2003A&A...407..485G}, it seems doubtful that the best fit slope for the
apertures in Figure 22 is indicative of a pattern speed.

\begin{figure}
\epsscale{.80}
\plotone{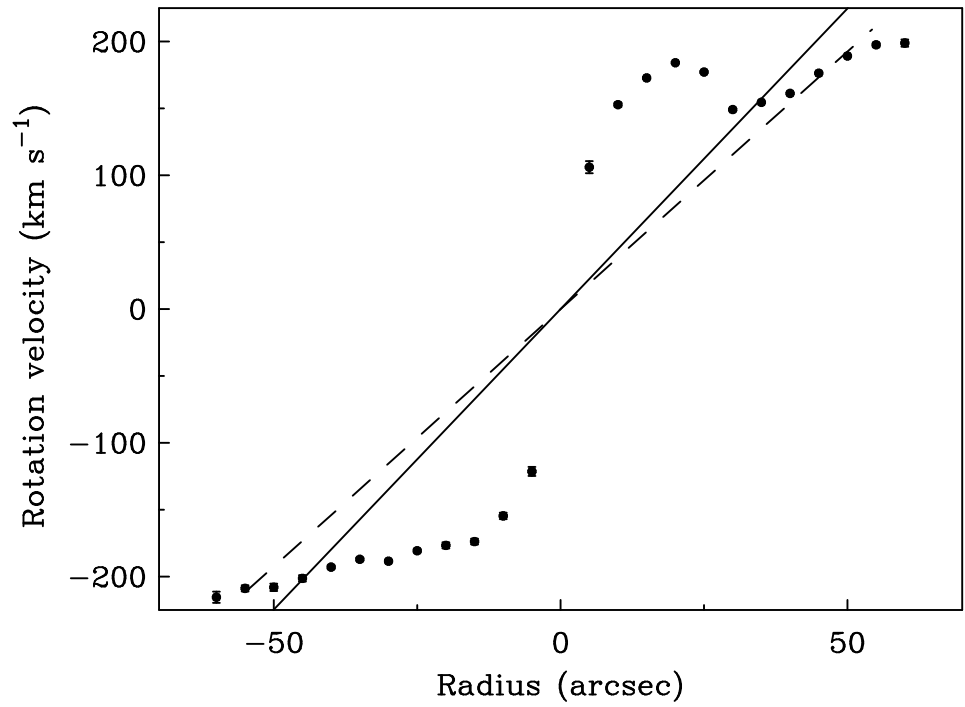}
\caption{Rotation curve for NGC 4826.  For most points, the formal error bars
are smaller than the plotting symbol size.  The solid and dashed lines
represent the best fit linear slopes to the rotation curve and the data in
Figure 22, respectively.
\label{fig23}}
\end{figure}

\section{Summary and Conclusions}

In this paper we have carried out experiments using a simulated barred galaxy
which have demonstrated the effectiveness of the TW method, and we have tested
the method on a clumpy, axisymmetric gas disk with no wave.  We have shown in
the former case that a correlation can naturally appear in
$<$$v$$>$--$<$$x$$>$ plots even when no large-scale wave is present.  Our bar
simulations showed that the TW method is robust, even at poor angular
resolution, except when the bar becomes too closely aligned with one of the
principle axes.

We have applied the method to high-resolution CO data cubes with full flux
information for six galaxies from the BIMA SONG survey, each of which has an
ISM dominated by molecular gas.  This work builds on the initial exploration
of this application of the TW method by ZRM.

For the two spiral galaxies with bars closely aligned with a principle axis,
NGC 3627 and NGC 4321, we find spiral pattern speeds in agreement with
previous determinations by other means.  For the two barred galaxies NGC 1068
and NGC 4736, where the bar is not closely aligned with either principle axis,
we have found some evidence that the bar and spiral patterns rotate with
different speeds, the same conclusion made for M51 by ZRM.  Unfortunately, the
nature of the method has made it difficult to determine unambiguously the bar
pattern speed in these cases because apertures which cross the bar also
include emission from the spiral.  Higher resolution observations would allow
more independent apertures to be placed across the bar region, reducing the
formal error in the best fit slope, but the above ambiguity would still
remain.  Nevertheless, the results demonstrate the power of the TW method to
address observationally the long standing question of how bar and spiral
pattern speeds relate.  Our results for the spiral patterns in these two
galaxies disagree with previous determinations based on identifying predicted
behaviors at resonance locations.  However, it is also possible to interpret
our results in such terms with some success, including tentative indications
of mode coupling.  Our results for NGC 4414 and NGC 4826, along with our tests
on a simulated clumpy disk, demonstrate that one must exercise caution in
applying the TW method to galaxies for which there is little evidence of a
large-scale wave.

As mentioned in the Introduction, further work will focus on galaxies
with HI-dominated ISMs and those with substantial amounts of both molecular
and atomic gas, from the BIMA SONG survey and elsewhere in the literature.

%% Included in this acknowledgments section are examples of the
%% AASTeX hypertext markup commands. Use \url without the optional [HREF]
%% argument when you want to print the url directly in the text. Otherwise,
%% use either \url or \anchor, with the HREF as the first argument and the
%% text to be printed in the second.

\acknowledgments

This material is based on work partially supported by the National Science
Foundation under Grant No. AST-0306958 to R.J.R.


%% copy editing.  Individual instruments can be provided in parentheses,
%% after the keyword, but they will not be verified.

Facilities: \facility{BIMA}.




%%\begin{thebibliography}{}
\bibliography{ms}
%%\end{thebibliography}

\clearpage

%% Tables should be submitted one per page, so put a \clearpage before

%% Two options are available to the author for producing tables:  the
%% deluxetable environment provided by the AASTeX package or the LaTeX
%% table environment.  Use of deluxetable is preferred.
%%

%% Three table samples follow, two marked up in the deluxetable environment,
%% one marked up as a LaTeX table.

\begin{deluxetable}{llccr}
\tabletypesize{\scriptsize}
\tablecaption{Galaxy Sample \label{tbl-1}}
\tablewidth{0pt}
\tablehead{
\colhead{Galaxy} & \colhead{Type}   & \colhead{Distance (Mpc)$^a$}   &
\colhead{$M_{\rm H_2}/M_{\rm HI}$$^b$} &
\colhead{Gas morphology comments}
}
\startdata
NGC 1068       & Sb    & 14.4 &   3.1 &  strong spiral/ring structure, bar  \\
NGC 3627       & SBb   & 11.1 &   3.0 &  strong bar and spiral             \\
NGC 4321 (M100)& SBbc   & 16.1 &   2.3 & strong spiral and bar      \\
NGC 4414       & Sc    & 19.1 &   1.2 &  flocculent spiral       \\
NGC 4736 (M94) & Sab   & 4.2  &   0.9 &  strong spiral and bar        \\
NGC 4826       & Sab   & 4.1  &   1.3 &  clumpy structure  \\

\enddata

%% Text for table notes should follow after the \enddata but before
%% the \end{deluxetable}. Make sure there is at least one \tablenotemark
%% in the table for each \tablenotetext.

\tablenotetext{a}{See references in text}
\tablenotetext{b}{From data compiled by \citet{1989ApJ...347L..55Y} and
\citet{1993A&A...272..123S}, assuming
$X = 2 \times 10^{20}\ {\rm mol\ cm^{-2}\ (K\ km\ s^{-1})^{-1}}.$}

\end{deluxetable}
\clearpage

\begin{deluxetable}{llccr}
\tabletypesize{\scriptsize}
\tablecaption{Galaxy Simulation Disk Parameters \label{tbl-2}}
\tablewidth{0pt}
\tablehead{
\colhead{Component} & \colhead{Mass}   & \colhead{Number of Particles}   &  fraction of particles \\
& \colhead{ Simulation Units} &  &\\
}
\startdata
Halo & 3.4 & 12800 & 0.4 \\
Bulge & 0.33 & 6400 & 0.2\\
Disk Stars & 3.06 & 6400 & 0.2 \\
Gas & 0.34 & 6400 & 0.2 \\
\enddata
 
%% Text for table notes should follow after the \enddata but before
%% the \end{deluxetable}. Make sure there is at least one \tablenotemark
%% in the table for each \tablenotetext.

\end{deluxetable}
\clearpage

%% The following command ends your manuscript. LaTeX will ignore any text

\end{document}